\documentclass[altaffilletter,aps,nofootinbib,twocolumn,prd,eqsecnum,preprintnumbers,superscriptaddress]{revtex4}
\pdfoutput=1
\usepackage[caption=false]{subfig}
\usepackage{graphicx}
\usepackage{amsmath}
\usepackage{amsfonts}
\usepackage{amssymb}
\usepackage{color}
\usepackage{bm}
\usepackage{mathrsfs}
\usepackage{epstopdf}
\usepackage{url}
\usepackage{footnote}
\usepackage{textcomp}
\usepackage{dsfont}
\usepackage{ulem}
\usepackage{hyperref}
\usepackage{enumerate}

\makeatletter
\newcommand*{\rom}[1]{\expandafter\@slowromancap\romannumeral #1@}
\makeatother

\begin{document}

\title{Probing Palatini-type gravity theories through gravitational wave detections via quasinormal modes}

\author{Che-Yu Chen}
\email{b97202056@gmail.com}
\affiliation{Department of Physics and Center for Theoretical Sciences, National Taiwan University, Taipei, Taiwan 10617}
\affiliation{LeCosPA, National Taiwan University, Taipei, Taiwan 10617}

\author{Mariam Bouhmadi-L\'{o}pez}
\email{mariam.bouhmadi@ehu.eus}
\affiliation{Department of Theoretical Physics, University of the Basque Country UPV/EHU, P.O.~Box 644, 48080 Bilbao, Spain}
\affiliation{IKERBASQUE, Basque Foundation for Science, 48011 Bilbao, Spain}

\author{Pisin Chen}
\email{pisinchen@phys.ntu.edu.tw}
\affiliation{Department of Physics and Center for Theoretical Sciences, National Taiwan University, Taipei, Taiwan 10617}
\affiliation{LeCosPA, National Taiwan University, Taipei, Taiwan 10617}
\affiliation{Kavli Institute for Particle Astrophysics and Cosmology, SLAC National Accelerator Laboratory, Stanford University, Stanford, CA 94305, USA}
\begin{abstract}
The possibility of testing gravity theories with the help of gravitational wave detections has become an interesting arena of recent research. In this paper, we follow this direction by investigating the quasinormal modes (QNMs) of the axial perturbations for charged black holes in the Palatini-type theories of gravity, specifically $(i)$ the Palatini $f(R)$ gravity coupled with Born-Infeld nonlinear electrodynamics and $(ii)$ the Eddington-inspired-Born-Infeld gravity (EiBI) coupled with Maxwell electromagnetic fields. The coupled master equations describing perturbations of charged black holes in these theories are obtained with the tetrad formalism. By using the Wentzel-Kramers-Brillouin (WKB) method up to 6th order, we calculate the QNM frequencies of the EiBI charged black holes, the Einstein-Born-Infeld black holes, and the Born-Infeld charged black holes within the Palatini $R+\alpha R^2$ gravity. The QNM spectra of these black holes would deviate from those of the Reissner-Nordstr\"om black hole. In addition, we study the QNMs in the eikonal limit and find that for the axial perturbations of the EiBI charged black holes, the link between the eikonal QNMs and the unstable null circular orbit around the black hole is violated.
\end{abstract}

\maketitle

\section{Introduction}   
It is quite fair to say that one of the most enchanting events of recent discovery in modern physics is the direct detection of gravitational waves (GWs) from the coalescence of binary black holes \cite{Abbott:2016blz,Abbott:2017oio}. The reason why the direct detections of GW signals are so appealing is that they not only confirm the predictions of Einstein's general relativity (GR) once again, but also render GWs spectacularly a suitable tool for human being to \textit{hear} deep into the sky far beyond the reach of electromagnetic signals. Not long after the first detection of GWs, the LIGO-VIRGO collaboration succeeded in detecting the GWs emitted from the merger of binary neutron stars, with an accurate localization of the source \cite{TheLIGOScientific:2017qsa}. The prompt and accompanied electromagnetic signals emitted from the source were also detected. This outstanding accomplishment has initiated a new era of multi-messenger astronomy. 

In addition, the direct detections of GWs could help us to test other gravitational theories, or even to falsify some extended theories of gravity \cite{Capozziello:2011et}. In fact, one of the reasons to consider extended theories of gravity is that GR inevitability predicts the existence of spacetime singularities like the big bang singularity and the black hole singularity. To ameliorate these spacetime singularities, one may resort to some extended theories of gravity which modify Einstein equation at the large curvature limits, but reduce to GR when curvature becomes small. Within the new era of GW astronomy, one plausible way to test these extended theories of gravity, for instance, is via the speed of GWs, as was done in Refs.~\cite{Lombriser:2015sxa,Lombriser:2016yzn,Sakstein:2017xjx,Baker:2017hug,Ezquiaga:2018btd}.

Another interesting aspect regarding GW detections could be the ringdown signals in the final stage of a merger event. Essentially, the final product of a merger event, no matter if seeded from binary black holes or from binary neutron stars, is usually a black hole. Before the final black hole settles itself, there is an intermediate stage where the distortion of the black hole is gradually relieved, with the emission of GWs. In practice, the ringdown stage can be described by the theory of black hole perturbations and the frequencies of the GWs are featured by quasinormal modes (QNMs). In this stage, the distorted black hole can be regarded as a dissipative system. The perturbations have a discrete spectrum and the QNM frequencies are complex numbers. The real part of the frequencies describes the oscillations of the perturbations and the imaginary part corresponds to the decay of the amplitude. Interestingly, these QNM frequencies merely depend on the parameters characterizing the black holes, such as the mass, the charge, and the spin. If there are additional parameters appearing in the underlying theory, these parameters should manifest themselves in the QNM spectra. Along this direction of research, the QNMs of black holes in several gravity theories have been investigated, such as in the Horndeski gravity \cite{Kobayashi:2012kh,Kobayashi:2014wsa,Minamitsuji:2014hha,Dong:2017toi,Tattersall:2018nve}, metric $f(R)$ gravity \cite{Sebastian:2014qra,Bhattacharyya:2017tyc,Bhattacharyya:2018qbe}, massive gravity \cite{Fernando:2014gda,Prasia:2016fcc}, Einstein-dilaton-Gauss-Bonnet gravity \cite{Blazquez-Salcedo:2016enn,Blazquez-Salcedo:2017txk,Blazquez-Salcedo:2018pxo,Blazquez-Salcedo:2016yka}, the Randall-Sundrum braneworld model \cite{Toshmatov:2016bsb}, Ho\v{r}ava-Lifshitz gravity \cite{Lin:2016wci}, higher dimensional black holes \cite{Cardoso:2003qd,Cardoso:2004cj,Cardoso:2003vt}, and Einstein-aether theory \cite{Ding:2017gfw}, etc. Furthermore, the QNMs of some regular black holes \cite{Flachi:2012nv,Toshmatov:2015wga} and the black holes with non-commutative geometry \cite{Liang:2018nmr,Das:2018fzc} have been analyzed in the literature. In addition, probing signatures of the black hole phase transitions in modified theories of gravity via QNMs has been shown to be possible \cite{Mahapatra:2016dae}. See Refs.~\cite{Nollert:1999ji,Berti:2009kk,Konoplya:2011qq,Berti:2015itd} for nice reviews on the latest progress of the field.

In this paper, as a further extension of our previous work \cite{Chen:2018mkf} in which the QNMs of massless scalar field perturbations were studied, we consider the QNMs of the axial perturbations for the charged black holes in two Palatini-type gravity theories: $(i)$ the Palatini $f(R)$ gravity coupled with Born-Infeld nonlinear electrodynamics (NED) and $(ii)$ the Eddington-inspired-Born-Infeld (EiBI) gravity coupled with linear electromagnetic fields. To calculate the QNM frequencies, we use the WKB method up to 6th order \cite{Schutz:1985zz,Iyer:1986np,Konoplya:2003ii,Matyjasek:2017psv}. We also calculate the QNMs in the eikonal limit in which the multipole number $l\rightarrow\infty$. Furthermore, the QNM frequencies will be compared with those of the Reissner-Nordstr\"om (RN) black hole in GR. Note as well that for the merger events of binary neutron stars, the ringdown timescale is usually shorter than the timescale of charge neutralization of the black hole \cite{Franklin:2011qn}. This justifies to some extent the validity of studying QNMs of charged black holes from the astrophysical point of view.

The charged black holes in the Palatini $f(R)$ gravity coupled with Born-Infeld NED have been studied in detail in Ref.~\cite{Olmo:2011ja}. The black hole solutions are very close to the RN black hole at the exterior spacetime, while deviate from it inside the event horizon because of the Born-Infeld NED source and the nonlinear function $f(R)$. It has been shown that there exist some regions in the parameter space where the singularity inside the event horizon is replaced with a finite size wormhole structure \cite{Bambi:2015zch}. Moreover, one can construct the Einstein-Born-Infeld (EBI) black hole by choosing $f(R)=R$. The properties of this charged black hole have been widely studied in the literature \cite{Breton:2001yk,Dey:2004yt,Cai:2004eh,Fernando:2003tz}. Again, due to the Born-Infeld corrections from the NED source, the interior structure of the black hole would change significantly as compared with that of the RN black hole in GR.

The EiBI gravity was formally proposed in Ref.~\cite{Banados:2010ix} to resolve the initial big bang singularity \cite{Scargill:2012kg}. This theory reduces to GR in vacuum but differs from it when matter is included. The exact expressions and some interesting properties of the charged black holes in the EiBI theory were studied in Refs.~\cite{Wei:2014dka,Sotani:2014lua} (see Ref.~\cite{BeltranJimenez:2017doy} for a review on the EiBI gravity). Due to the Born-Infeld corrections from the gravity sector, the interior structure of the black hole could deviate from that of the RN black hole notably \cite{Olmo:2013gqa,Olmo:2015bya,Olmo:2015dba}. One can then compare the QNM frequencies of the EBI black holes and those of the EiBI charged black holes to see how the Born-Infeld structure from the matter and the gravitational sector affects the QNMs.

This paper is outlined as follows. In section \ref{secII}, we briefly review the tetrad formalism which will be used later to derive the master equations describing the axial perturbations of the black holes. In section \ref{sectIII}, the perturbed Maxwell equation for NED is obtained for the sake of later convenience. In section \ref{secIV}, we derive sequentially the coupled master equations of the axial perturbations for charged black holes in the Palatini $f(R)$ gravity coupled with Born-Infeld NED and in the EiBI gravity with linear electromagnetic fields. In section \ref{sectV}, we calculate the QNM frequencies of these charged black holes by using the WKB semi-analytic method. The QNMs in the eikonal limit are discussed as well. We finally conclude in section \ref{conclu}.

\section{Tetrad formalism}\label{secII}
To study the QNMs of the black holes of our interest, we shall consider the perturbations of a static and spherically symmetric spacetime. Without loss of generality, the perturbed spacetime can be described by a non-stationary and axisymmetric metric in which the symmetrical axis is turned in such a way that no $\phi$ dependence appears in the metric functions. In general, the metric can be written as follows \cite{Chandrabook}:
\begin{align}
ds^2=&-e^{2\nu}\left(dx^0\right)^2+e^{2\psi}\left(dx^1-\sigma dx^0-q_2dx^2-q_3dx^3\right)^2\nonumber\\&+e^{2\mu_2}\left(dx^2\right)^2+e^{2\mu_3}\left(dx^3\right)^2\,,\label{metricg}
\end{align}
where $\nu$, $\psi$, $\mu_2$, $\mu_3$, $\sigma$, $q_2$, and $q_3$ are functions of time $t$ ($t=x^0$), radial coordinate $r$ ($r=x^2$), and polar angle $\theta$ ($\theta=x^3$). Because the system is axisymmetric, the metric functions are independent of the azimuthal angle $\phi$ ($\phi=x^1$). In this work, the notation used in Ref.~\cite{Chandrabook} is strictly followed. The only difference is that the metric function $\omega$ used in Ref.~\cite{Chandrabook} is replaced with $\sigma$ in Eq.~\eqref{metricg} since we will use $\omega$ to denote the frequency of the perturbations later. Note that in the background spacetime which is static and spherically symmetric, we have $\sigma=q_2=q_3=0$.

To study the perturbations of the spacetime metric \eqref{metricg}, we will use the tetrad formalism in which one defines a basis associated with the metric \eqref{metricg} \cite{Chandrabook}:
\begin{align}
e^{\mu}_{(0)}&=\left(e^{-\nu},\quad\sigma e^{-\nu},\quad0,\quad0\right)\,,\nonumber\\
e^{\mu}_{(1)}&=\left(0,\quad e^{-\psi},\quad 0,\quad0\right)\,,\nonumber\\
e^{\mu}_{(2)}&=\left(0,\quad q_2e^{-\mu_2},\quad e^{-\mu_2},\quad0\right)\,,\nonumber\\
e^{\mu}_{(3)}&=\left(0,\quad q_3e^{-\mu_3},\quad 0,\quad e^{-\mu_3}\right)\,,\label{tetradbasis111}
\end{align}
and
\begin{align}
e_{\mu}^{(0)}&=\left(e^{\nu},\quad0,\quad0,\quad0\right)\,,\nonumber\\
e_{\mu}^{(1)}&=\left(-\sigma e^{\psi},\quad e^{\psi},\quad -q_2e^{\psi},\quad -q_3e^{\psi}\right)\,,\nonumber\\
e_{\mu}^{(2)}&=\left(0,\quad 0,\quad e^{\mu_2},\quad0\right)\,,\nonumber\\
e_{\mu}^{(3)}&=\left(0,\quad 0,\quad 0,\quad e^{\mu_3}\right)\,,\label{tetradbasis222}
\end{align}
where the tetrad indices are enclosed in parentheses to distinguish them from the tensor indices. The tetrad basis should satisfy
\begin{align}
e_{\mu}^{(a)}e^{\mu}_{(b)}&=\delta^{(a)}_{(b)}\,,\quad e_{\mu}^{(a)}e^{\nu}_{(a)}=\delta^{\nu}_{\mu}\,,\nonumber\\
e_{\mu}^{(a)}&=g_{\mu\nu}\eta^{(a)(b)}e^{\nu}_{(b)}\,,\nonumber\\
g_{\mu\nu}&=\eta_{(a)(b)}e_{\mu}^{(a)}e_{\nu}^{(b)}\equiv e_{(a)\mu}e_{\nu}^{(a)}\,.
\end{align}
Conceptually, in the tetrad formalism we project the relevant quantities defined on the coordinate basis of $g_{\mu\nu}$ onto a chosen basis of $\eta_{(a)(b)}$ by constructing the tetrad basis correspondingly. In practice, $\eta_{(a)(b)}$ is usually assumed to be the Minkowskian matrix
\begin{equation}
\eta_{(a)(b)}=\eta^{(a)(b)}=\textrm{diag}\left(-1,1,1,1\right)\,.
\end{equation}
In this regard, any vector or tensor field can be projected onto the tetrad frame in which the field can be expressed through its tetrad components:
\begin{align}
A_{\mu}&=e_{\mu}^{(a)}A_{(a)}\,,\quad A_{(a)}=e_{(a)}^{\mu}A_{\mu}\,,\nonumber\\
B_{\mu\nu}&=e_{\mu}^{(a)}e_{\nu}^{(b)}B_{(a)(b)}\,,\quad B_{(a)(b)}=e_{(a)}^{\mu}e_{(b)}^{\nu}B_{\mu\nu}\,.
\end{align}

It has been shown in Ref.~\cite{Chandrabook} that the master equations describing the gravitational perturbations of black holes (Schwarzschild, RN, etc) can be obtained by using the tetrad formalism in a straightforward and concise manner. One should notice that in the tetrad formalism, the covariant (partial) derivative in the original coordinate frame is replaced with the intrinsic (directional) derivative in the tetrad frame. For instance, the derivatives of an arbitrary rank two object $H_{\mu\nu}$ in the two frames can be related as follows \cite{Chandrabook}
\begin{align}
&\,H_{(a)(b)|(c)}\equiv e^{\lambda}_{(c)}H_{\mu\nu;\lambda}e_{(a)}^{\mu}e_{(b)}^{\nu}\nonumber\\
=&\,H_{(a)(b),(c)}\nonumber\\&-\eta^{(m)(n)}\left(\gamma_{(n)(a)(c)}H_{(m)(b)}+\gamma_{(n)(b)(c)}H_{(a)(m)}\right)\,,\label{2.7}
\end{align}
where a vertical rule and a comma denote the intrinsic and directional derivative with respect to the tetrad indices, respectively. A semicolon denotes a covariant derivative with respect to the tensor indices. Furthermore, the Ricci rotation coefficients are defined by
\begin{equation}
\gamma_{(c)(a)(b)}\equiv e_{(b)}^{\mu}e_{(a)\nu;\mu}e_{(c)}^{\nu}\,,
\end{equation}
and their components corresponding to the metric \eqref{metricg} are given in Ref.~\cite{Chandrabook}.

\section{Perturbed Maxwell equation for NED}\label{sectIII}
In a gravity theory formulated upon the Palatini variational principle, the matter Lagrangian is assumed to be coupled with the physical metric $g_{\mu\nu}$ only. Therefore, the matter fields would follow the geodesics defined by this metric and the conservation equation of the energy momentum tensor follows the standard form with respect to $g_{\mu\nu}$. In this section, we focus on the matter Lagrangian described by NED \cite{Olmo:2011ja}:
\begin{equation}
\mathcal{S}_m=\frac{1}{8\pi}\int d^4x\sqrt{-g}\phi(X,Y)\,,\label{matterNEDLA}
\end{equation}
where we have set $G=c=1$, and $\phi(X,Y)$ is a function of gauge field invariants defined by \cite{Olmo:2011ja}:
\begin{equation}
X\equiv-\frac{1}{2}F_{\mu\nu}F^{\mu\nu}\,,\qquad Y\equiv-\frac{1}{2}F_{\mu\nu}F^{*\mu\nu}\,,
\end{equation}
where $F^{*\mu\nu}\equiv\frac{1}{2}\epsilon^{\mu\nu\alpha\beta}F_{\alpha\beta}$ is the dual of the field strength. The standard Maxwell electromagnetic fields are recovered when $\phi(X,Y)=X$; For the sake of simplicity, we will assume a vanishing magnetic field, i.e., $Y=0$, in the rest of this paper.

For a gravitational theory minimally coupled to NED with a purely radial electric field and no magnetic field, only the $(t,r)$ and $(r,t)$ components, i.e. the $(0,2)$ and $(2,0)$ components of the field strength $F_{\mu\nu}$ appear at the background level. In the tetrad frame, the field strength $F_{(a)(b)}$ at the background level satisfies \cite{Olmo:2011ja}
\begin{equation}
F_{02}=F_{(0)(2)}e_0^{(0)}e_2^{(2)}=F_{(0)(2)}e^{\nu+\mu_2}=\frac{Q_*e^{\nu+\mu_2}}{r^2\phi_X}\,,\label{backgrondF}
\end{equation}
where $\phi_X=d\phi/dX$ and $Q_*$ is an integration constant which can be regarded as the charge of the black hole. Note that the last equality in Eq.~\eqref{backgrondF} can be obtained from the conservation equation of NED at the background level \cite{Olmo:2011ja}. From Eq.~\eqref{backgrondF}, we get
\begin{equation}
F_{(0)(2)}=\frac{Q_*}{r^2\phi_X}\,.\qquad\textrm{(background level)}\label{backgroundFF}
\end{equation}

In the general case where the perturbations are taken into account, the metric functions and the field strength could depend on $t$, $r$, and $\theta$. In this case, the Bianchi identity of the field strength $F_{[(a)(b)|(c)]}=0$ leads to
\begin{align}
&\left(e^{\psi+\mu_2}F_{(1)(2)}\right)_{,3}+\left(e^{\psi+\mu_3}F_{(3)(1)}\right)_{,2}=0\,,\label{omit1}\\
&\left(e^{\psi+\nu}F_{(0)(1)}\right)_{,2}+\left(e^{\psi+\mu_2}F_{(1)(2)}\right)_{,0}=0\,,\label{3.6}\\
&\left(e^{\psi+\nu}F_{(0)(1)}\right)_{,3}+\left(e^{\psi+\mu_3}F_{(1)(3)}\right)_{,0}=0\,,\label{3.7}\\
&\left(e^{\mu_2+\nu}F_{(0)(2)}\right)_{,3}+\left(e^{\mu_3+\nu}F_{(3)(0)}\right)_{,2}+\left(e^{\mu_2+\mu_3}F_{(2)(3)}\right)_{,0}\nonumber\\
=&-e^{\psi+\nu}\left(q_{3,2}-q_{2,3}\right)F_{(0)(1)}+e^{\psi+\mu_2}\left(\sigma_{,3}-q_{3,0}\right)F_{(1)(2)}\nonumber\\&+e^{\psi+\mu_3}\left(\sigma_{,2}-q_{2,0}\right)F_{(3)(1)}\,.\label{3.8}
\end{align}
Note that the comma here denotes the partial derivative with respect to the tensor indices. This derivative is related to the directional derivative shown in Eq.~\eqref{2.7} by $H_{(a)(b),(c)}=e^{\mu}_{(c)}H_{(a)(b),\mu}$ \cite{Chandrabook}. In addition, Eq.~\eqref{omit1} is a redundant equation because it is just an integrability condition for Eqs.~\eqref{3.6} and \eqref{3.7}. 

On the other hand, the conservation equation for NED $\eta^{(b)(c)}(F_{(a)(b)}\phi_X)_{|(c)}=0$ can be written explicitly as follows
\begin{align}
&\left(e^{\psi+\mu_3}F_{(0)(2)}\phi_X\right)_{,2}+\left(e^{\psi+\mu_2}F_{(0)(3)}\phi_X\right)_{,3}=0\,,\label{omit2}\\
&\left(e^{\psi+\mu_2}F_{(0)(3)}\phi_X\right)_{,0}+\left(e^{\psi+\nu}F_{(3)(2)}\phi_X\right)_{,2}=0\,,\label{3.10}\\
&\left(e^{\psi+\nu}F_{(2)(3)}\phi_X\right)_{,3}+\left(e^{\psi+\mu_3}F_{(0)(2)}\phi_X\right)_{,0}=0\,,\label{3.11}\\
&\left(e^{\nu+\mu_3}F_{(1)(2)}\phi_X\right)_{,2}+\left(e^{\nu+\mu_2}F_{(1)(3)}\phi_X\right)_{,3}\nonumber\\&+\left(e^{\mu_2+\mu_3}F_{(0)(1)}\phi_X\right)_{,0}\nonumber\\=&\,\Big[e^{\psi+\mu_3}\left(\sigma_{,2}-q_{2,0}\right)F_{(0)(2)}+e^{\psi+\mu_2}\left(\sigma_{,3}-q_{3,0}\right)F_{(0)(3)}\nonumber\\&+e^{\psi+\nu}\left(q_{3,2}-q_{2,3}\right)F_{(2)(3)}\Big]\phi_X\,.\label{3.12}
\end{align}
Again, it can be shown that Eq.~\eqref{omit2} is a redundant equation since it is an integrability condition for Eqs.~\eqref{3.10} and \eqref{3.11}. 

To linearize the equations above, the scalar field $\phi_X$ and the gauge field invariant $X$ should be decomposed into the background and the 1st order parts:
\begin{equation}
\phi_X\rightarrow\phi_{X}+\delta\phi_{X}\,,\quad X\rightarrow X+\delta X\,,
\end{equation}
where
\begin{equation}
\delta\phi_{X}=\phi_{XX}\delta X=2\phi_{XX}F_{(0)(2)}\delta F_{(0)(2)}\,.\label{3.14}
\end{equation}
In the above expressions, we have decomposed $F_{(0)(2)}$ as $F_{(0)(2)}\rightarrow F_{(0)(2)}+\delta F_{(0)(2)}$ in which the background $F_{(0)(2)}$ is given in{\footnote{We only use a delta into the linear order perturbations of quantities whose values at the background level do not vanish. The quantities that vanish at the background level, such as the metric functions $\sigma$, $q_2$, and $q_3$, and the Maxwell tensor components $F_{(i)(j)}$ ($ij\ne02$ or $20$), shall be regarded as linear order perturbation quantities directly.}} Eq.~\eqref{backgroundFF}. Note that the quantity $X$ at the background level is given by $X=F_{(0)(2)}^2$.

After the decomposition, the linearized Eqs.~\eqref{3.6}, \eqref{3.7}, and \eqref{3.8} are
\begin{align}
\left(re^\nu\sin{\theta} F_{(0)(1)}\right)_{,r}+re^{\mu_2}\sin{\theta}F_{(1)(2),0}&=0\,,\label{12}\\
re^\nu\left(F_{(0)(1)}\sin{\theta}\right)_{,\theta}+r^2\sin{\theta}F_{(1)(3),0}&=0\,,\label{13}\\
e^{\mu_2+\nu}\left[\delta F_{(0)(2),\theta}+F_{(0)(2)}\left(\delta\mu_2+\delta\nu\right)_{,\theta}\right]\nonumber\\+\left(re^\nu F_{(3)(0)}\right)_{,r}+re^{\mu_2}F_{(2)(3),0}&=0\,.
\end{align}
On the other hand, the linearized Eqs.~\eqref{3.10}, \eqref{3.11} and \eqref{3.12} can be written as
\begin{align}
&\phi_{X}re^{\mu_2}F_{(0)(3),0}+\left(\phi_{X}re^\nu F_{(3)(2)}\right)_{,r}=0\,,\\
&\phi_{X}\left[\delta F_{(0)(2),0}+F_{(0)(2)}\left(\delta\psi+\delta\mu_3\right)_{,0}\right]\nonumber\\+&\,\frac{\phi_{X}e^{\nu}}{r\sin{\theta}}\left(F_{(2)(3)}\sin{\theta}\right)_{,\theta}+\delta\phi_{X,0}F_{(0)(2)}=0\,,\\
&\left(re^\nu\phi_{X}F_{(1)(2)}\right)_{,r}+e^{\nu+\mu_2}\phi_{X}F_{(1)(3),\theta}\nonumber\\+&\,re^{\mu_2}\phi_{X}F_{(0)(1),0}=r^2\sin{\theta}F_{(0)(2)}\phi_{X}\left(\sigma_{,2}-q_{2,0}\right)\,,\label{17}
\end{align}
where $\delta\nu$, $\delta\mu_2$, $\delta\mu_3$, and $\delta\psi$ are the 1st order parts of the metric functions $\nu$, $\mu_2$, $\mu_3$, and $\psi$, respectively.

For the sake of abbreviation, we define the field perturbation
\begin{equation}
B\equiv F_{(0)(1)}\sin{\theta}\,.
\end{equation}
After differentiating Eq.~\eqref{17} with respect to $x^0=t$ and using Eqs.~\eqref{12} and \eqref{13}, we have
\begin{align}
&\left[\phi_{X}e^{\nu-\mu_2}\left(re^\nu B\right)_{,r}\right]_{,r}+\phi_{X}\frac{e^{2\nu+\mu_2}}{r}\left(\frac{B_{,\theta}}{\sin{\theta}}\right)_{,\theta}\sin{\theta}\nonumber\\-&\,\phi_{X}re^{\mu_2}B_{,00}
=-r^2F_{(0)(2)}\phi_{X}\left(\sigma_{,20}-q_{2,00}\right)\sin^2{\theta}\,.\label{21}
\end{align}
Recall that in the derivation of Eq.~\eqref{21}, we have only used the Maxwell equation of the NED source. Eq.~\eqref{21} would become one of the coupled master equations in the sense that the linear order metric functions on the right hand side of Eq.~\eqref{21} are coupled with the field perturbation $B$ (or the 1st order field strength $F_{(0)(1)}$) on the left hand side. In order to deduce the other coupled equation, the perturbed gravitational equation should be taken into account.

At the end of this section, we would like to write down the perturbed energy momentum tensor of NED, which will appear on the right hand side of the perturbed gravitational equations later. The energy momentum tensor of NED in the tetrad frame is
\begin{equation}
T_{(a)(b)}=\frac{1}{4\pi}\left(\phi_X {F_{(a)}}^{(m)}F_{(b)(m)}+\frac{1}{2}\phi\eta_{(a)(b)}\right)\,.
\end{equation}
Furthermore, the perturbed energy momentum tensor reads
\begin{align}
\delta T_{(a)(b)}=&\,\frac{1}{4\pi}\Big(\phi_X {\delta F_{(a)}}^{(m)}F_{(b)(m)}+\phi_X {F_{(a)}}^{(m)}\delta F_{(b)(m)}\nonumber\\&+\delta\phi_{X}{F_{(a)}}^{(m)}F_{(b)(m)}+\frac{1}{2}\eta_{(a)(b)}\phi_X\delta X\Big)\,.
\end{align}
Finally, we can write down its components explicitly as follows:
\begin{align}
\delta T_{(0)(0)}&=-\delta T_{(2)(2)}=\frac{1}{4\pi}\left(\phi_X+2\phi_{XX}X\right)F_{(0)(2)}\delta F_{(0)(2)}\,,\\
\delta T_{(3)(3)}&=\delta T_{(1)(1)}=\frac{1}{4\pi}\phi_XF_{(0)(2)}\delta F_{(0)(2)}\,,\\
\delta T_{(0)(1)}&=\frac{1}{4\pi}\phi_XF_{(0)(2)}F_{(1)(2)}\,,\nonumber\\\delta T_{(0)(3)}&=-\frac{1}{4\pi}\phi_XF_{(0)(2)}F_{(2)(3)}\,,\nonumber\\
\delta T_{(1)(2)}&=-\frac{1}{4\pi}\phi_XF_{(0)(2)}F_{(0)(1)}\,,\nonumber\\\delta T_{(2)(3)}&=-\frac{1}{4\pi}\phi_XF_{(0)(2)}F_{(0)(3)}\,,\\
\delta T_{(0)(2)}&=\delta T_{(1)(3)}=0\,.
\end{align}

\section{The master equations}\label{secIV}
As we have mentioned previously, the master equations describing the gravitational perturbations of a charged black hole are two coupled equations. This is because of the coupling between the gravitational field and the electromagnetic field in the system. So far we have derived one of the coupled equations, i.e., Eq.~\eqref{21}, from the Maxwell equation of the NED source. In this section, we will carry out the derivation of the other coupled equation from the gravitational equation of the theory. We will first consider the Palatini $f(R)$ gravity coupled with NED and obtain the master equations in this theory. After that we will turn to deduce the master equations of the EiBI gravity coupled with linear electromagnetic fields.

\subsection{Axial perturbations of charged black holes in Palatini $f(R)$ with NED}
In this subsection, we will consider the Palatini $f(R)$ theory coupled with NED. The action of the theory reads \cite{Olmo:2011ja}
\begin{equation}
\mathcal{S}_1=\frac{1}{16\pi}\int d^4x\sqrt{-g}f(R)+\mathcal{S}_m\,,
\end{equation}
where the matter Lagrangian $\mathcal{S}_m$ is given in Eq.~\eqref{matterNEDLA}. We would like to emphasize again that the theory is formulated within the Palatini variational principle in which the metric $g_{\mu\nu}$ and the affine connection $\Gamma$ are independent variables. For a nonlinear function $f$ of the Ricci scalar $R\equiv g^{\mu\nu} R_{\mu\nu}(\Gamma)$, the equations of motion would be different from those in the metric $f(R)$ theory. 

In addition, after deriving the master equations we will consider a particular NED model, that is, the Born-Infeld NED:
\begin{equation}
\phi(X)=2\beta_m^2\left(1-\sqrt{1-\frac{X}{\beta_m^2}}\right)\,.\label{BINEDla}
\end{equation}
The background solutions of the charged black holes in the Palatini $f(R)$ gravity coupled with the Born-Infeld NED Lagrangian \eqref{BINEDla} have been studied in Ref.~\cite{Olmo:2011ja}. The QNMs of a massless scalar field of such black holes have been discussed in Ref.~\cite{Chen:2018mkf}. The most general form of the metric functions of these black holes have been derived in Ref.~\cite{Olmo:2011ja} as well (see also Eqs.~(3.24) and (3.25) in Ref.~\cite{Chen:2018mkf}). 

Specifically, if we focus on Einstein gravity ($f(R)=R$) coupled with the Born-Infeld NED, we would get the Einstein-Born-Infeld (EBI) black hole whose deviations from the RN black hole result purely from the matter sector. Its exact metric functions at the background level read \cite{Chen:2018mkf,Breton:2001yk,Dey:2004yt,Cai:2004eh,Fernando:2003tz}
\begin{widetext}
\begin{equation}
e^{2\nu}=e^{-2\mu_2}=1-\frac{1}{r}-\frac{2\beta_m^2}{3}\left[\sqrt{r^4+r_m^4}-r^2-\frac{2r_m^4}{r^2}F\left(\frac{1}{4},\frac{1}{2};\frac{5}{4};-\frac{r_m^4}{r^4}\right)\right]\,,\quad e^{2\mu_3}=r^2\,,\quad e^{2\psi}=r^2\sin^2\theta\,,\label{EBIpsif}
\end{equation}
\end{widetext}
where $F(..,..;..;..)$ is the hypergeometric function \cite{Abramow} and $r_m$ is defined as $r_m\equiv\sqrt{Q_*/\beta_m}$. Note that we have used the following dimensionless rescalings:
\begin{equation}
\frac{Q_*}{r_s}\rightarrow Q_*\qquad \beta_mr_s\rightarrow \beta_m\qquad\frac{r}{r_s}\rightarrow r\,,
\end{equation} 
where $r_s/2\equiv M_0$ denotes the mass of the black hole seen by an observer infinitely far away. For the sake of convenience, we will use these rescalings in the rest of this paper.

As mentioned in Ref.~\cite{Chen:2018mkf}, it is interesting to compare the QNMs of the EBI charged black hole with those of the charged black holes within the EiBI gravity coupled with Maxwell electromagnetic fields. One can then compare directly the QNMs of charged black holes within two theories, one with the Born-Infeld correction from the matter sector and the other one with this kind of modification from the gravitational sector.

\subsubsection{Perturbed field equations}
In a gravitational theory constructed on the Palatini variational principle, the variation of the gravitational action with respect to the affine connection determines the auxiliary metric $q_{\mu\nu}$ which is compatible with the affine connection:
\begin{align}
ds_q^2=&-e^{2\tilde{\nu}}\left(dx^0\right)^2+e^{2\tilde{\psi}}\left(dx^1-\tilde{\sigma} dx^0-\tilde{q}_2dx^2-\tilde{q}_3dx^3\right)^2\nonumber\\&+e^{2\tilde{\mu}_2}\left(dx^2\right)^2+e^{2\tilde{\mu}_3}\left(dx^3\right)^2\,.\label{qmetric}
\end{align}
In the Palatini $f(R)$ gravity, the auxiliary metric and the physical metric are conformally related $q_{\mu\nu}=f_Rg_{\mu\nu}$, where $f_R=df/dR$. Therefore, their metric functions are related as follows
\begin{align}
e^{2\tilde{\nu}}&=f_Re^{2\nu}\,,\qquad e^{2\tilde{\psi}}=f_Re^{2\psi}\,,\nonumber\\ e^{2\tilde{\mu}_2}&=f_Re^{2\mu_2}\,,\qquad e^{2\tilde{\mu}_3}=f_Re^{2\mu_3}\,,\nonumber\\
\tilde\sigma&=\sigma\,,\quad\tilde{q}_2=q_2\,,\quad\tilde{q}_3=q_3\,.
\end{align}
On the other hand, the variation of the action with respect to $g_{\mu\nu}$ reads
\begin{equation}
f_RR_{(\mu\nu)}(q)-\frac{1}{2}g_{\mu\nu}f(R)=8\pi T_{\mu\nu}\,.\label{fR1nomal}
\end{equation}
It should be emphasized that the symmetric part of the Ricci tensor $R_{(\mu\nu)}$ in Eq.~\eqref{fR1nomal} is defined solely by the affine connection $\Gamma$. Because the field equation ensures the compatibility between the auxiliary metric $q_{\mu\nu}$ and the affine connection, it is fair to say that the Ricci tensor is defined by the auxiliary metric, that is, $R_{(\mu\nu)}(q)$.

In order to recast Eq.~\eqref{fR1nomal} into the tetrad frame, we rewrite Eq.~\eqref{fR1nomal} as follows
\begin{equation}
f_R{\tilde{e}^{(a)}}_{\mu}{\tilde{e}^{(b)}}_{\nu}\tilde{R}_{(a)(b)}-\frac{1}{2}fe^{(a)}_{\mu}e^{(b)}_{\nu}\eta_{(a)(b)}=8\pi e^{(a)}_{\mu}e^{(b)}_{\nu}T_{(a)(b)}\,,
\end{equation}
where $\tilde{R}_{(a)(b)}$ satisfies $R_{\mu\nu}(q)=\tilde{R}_{(a)(b)}{\tilde{e}^{(a)}}_{\mu}{\tilde{e}^{(b)}}_{\nu}$, and ${\tilde{e}^{(a)}}_{\mu}$ is a tetrad basis mapping the auxiliary metric $q_{\mu\nu}$ onto the tetrad frame. More explicitly, it satisfies
\begin{equation}
q_{\mu\nu}=\eta_{(a)(b)}{\tilde{e}^{(a)}}_{\mu}{\tilde{e}^{(b)}}_{\nu}\,.
\end{equation}
This additional tetrad basis is related to the physical tetrad basis according to the conformal relation between the two metrics:
\begin{equation}
\eta_{(a)(b)}{\tilde{e}^{(a)}}_{\mu}{\tilde{e}^{(b)}}_{\nu}=f_R\eta_{(a)(b)}e^{(a)}_{\mu}e^{(b)}_{\nu}\,.\label{eefRrelation}
\end{equation}
In this regard, Eq.~\eqref{fR1nomal} can be rewritten as
\begin{equation}
f_R^2\tilde{R}_{(a)(b)}-\frac{1}{2}f\eta_{(a)(b)}=8\pi f_RT_{(c)(d)}e^{(c)}_{\mu}e^{(d)}_{\nu}\tilde{e}^{\mu}_{(a)}\tilde{e}^{\nu}_{(b)}\,,\label{fR1}
\end{equation}
where the expressions of $\tilde{R}_{(a)(b)}$ are given in Ref.~\cite{Chandrabook}, in which all quantities should be replaced with their tilde counterparts. Furthermore, the scalar curvature $R$ can be written as
\begin{equation}
R= g^{\mu\nu}R_{\mu\nu}(q)=\eta^{(a)(b)}\tilde{R}_{(a)(b)}f_R\,.
\end{equation}

Since we are focusing on the axial perturbations (odd parity perturbations) which change sign when $\phi\rightarrow-\phi$, we only consider the $(1,3)$, $(1,2)$ and $(0,1)$ components of the linearized Eq.~\eqref{fR1}:
\begin{align}
&\left[f_Rr^2e^{\nu-\mu_2}\left(q_{2,3}-q_{3,2}\right)\right]_{,2}-f_Rr^2e^{-\nu+\mu_2}\left(\sigma_{,3}-q_{3,0}\right)_{,0}\nonumber\\=&\,0\,,\label{35}\\
&\left[f_Rr^2e^{\nu-\mu_2}\left(q_{3,2}-q_{2,3}\right)\sin^3{\theta}\right]_{,3}\nonumber\\&-\,f_Rr^4e^{-\nu-\mu_2}\left(\sigma_{,2}-q_{2,0}\right)_{,0}\sin^3{\theta}\nonumber\\
=&\,4\phi_XF_{(0)(2)}r^3e^\nu B\sin{\theta}\,,\label{36}\\
&\left[f_Rr^4e^{-\nu-\mu_2}\left(\sigma_{,2}-q_{2,0}\right)\sin^3{\theta}\right]_{,2}\nonumber\\&+f_Rr^2e^{-\nu+\mu_2}\left[\left(\sigma_{,3}-q_{3,0}\right)\sin^3{\theta}\right]_{,3}\nonumber\\
=&\,4\phi_XF_{(0)(2)}r^3e^{\mu_2}\sin^2{\theta}F_{(1)(2)}\,.\label{37}
\end{align}
Then, we define
\begin{equation}
Q\equiv f_Rr^2e^{\nu-\mu_2}\left(q_{2,3}-q_{3,2}\right)\sin^3{\theta}\,,
\end{equation}
with which Eqs.~\eqref{35} and \eqref{36} can be rewritten as
\begin{align}
e^{\nu-\mu_2}\frac{Q_{,2}}{f_Rr^2\sin^3{\theta}}&=\left(\sigma_{,3}-q_{3,0}\right)_{,0}\,,\label{4022}\\
e^{\nu+\mu_2}\frac{Q_{,3}}{f_Rr^4\sin^3{\theta}}&=-\left(\sigma_{,2}-q_{2,0}\right)_{,0}-e^{2\nu+\mu_2}\frac{4\phi_XF_{(0)(2)}}{f_Rr\sin^2{\theta}}B\,.\label{40}
\end{align}
By differentiating Eqs.~\eqref{4022} and \eqref{40} and eliminating $\sigma$, we obtain
\begin{align}
&\frac{1}{\sin^3{\theta}}\left(\frac{e^{\nu-\mu_2}}{f_Rr^2}Q_{,2}\right)_{,2}+\frac{e^{\nu+\mu_2}}{f_Rr^4}\left(\frac{Q_{,3}}{\sin^3{\theta}}\right)_{,3}\nonumber\\
=&\,\frac{Q_{,00}}{f_Rr^2\sin^3{\theta}e^{\nu-\mu_2}}-\frac{4e^{2\nu+\mu_2}\phi_XF_{(0)(2)}}{f_Rr}\left(\frac{B}{\sin^2\theta}\right)_{,3}\,.\label{41w}
\end{align}
Eq.~\eqref{41w} is the second equation of the coupled master equations describing the axial perturbations of a charged black hole within the Palatini $f(R)$ gravity coupled with NED. 

Finally, the term $(\sigma_{,20}-q_{2,00})$ in Eq.~\eqref{21} can be eliminated by using Eq.~\eqref{40}:
\begin{align}
&\left[\phi_Xe^{\nu-\mu_2}\left(re^\nu B\right)_{,2}\right]_{,2}+\phi_X\frac{e^{2\nu+\mu_2}}{r}\left(\frac{B_{,3}}{\sin{\theta}}\right)_{,3}\sin{\theta}\nonumber\\&-\phi_Xre^{\mu_2}B_{,00}\nonumber\\
=&\,4re^{2\nu+\mu_2}\left(\phi_XF_{(0)(2)}\right)^2\frac{B}{f_R}+e^{\nu+\mu_2}\frac{\phi_XF_{(0)(2)}}{f_Rr^2\sin{\theta}}Q_{,3}\,.\label{42w}
\end{align}
Eqs.~\eqref{41w} and \eqref{42w} form the coupled master equations describing the evolutions of the perturbation fields $B$ and $Q$, which correspond to the perturbations of the matter field and the gravitational field, respectively.

\subsubsection{Effective potentials}
For the sake of later convenience, we would like to recast the coupled master equations \eqref{41w} and \eqref{42w} into a Schr\"odinger-like form, which in practice is more suitable for the calculations of QNMs with the WKB method. We consider the ansatz \cite{Chandrabook}
\begin{equation}
Q(r,\theta)=Q(r)Y(\theta)\,,\qquad B(r,\theta)=B(r)Y_{,\theta}/\sin{\theta}\,,\label{43}
\end{equation}
where $Y(\theta)$ is the Gegenbauer function satisfying \cite{Abramow}
\begin{equation}
\frac{d}{d\theta}\left(\frac{1}{\sin^3{\theta}}\frac{dY}{d\theta}\right)=-\mu^2\frac{Y}{\sin^3{\theta}}\,,\label{gegen1}
\end{equation}
where $\mu^2=(l-1)(l+2)$ and $l$ is the multipole number. From Eq.~\eqref{gegen1}, it can be proven that
\begin{equation}
\sin{\theta}\frac{d}{d\theta}\left(\frac{1}{\sin{\theta}}\frac{d}{d\theta}\frac{Y_{,\theta}}{\sin{\theta}}\right)=-\left(\mu^2+2\right)\frac{Y_{,\theta}}{\sin{\theta}}\,.\label{gegen2}
\end{equation}
With the assumption \eqref{43}, the coupled Eqs.~\eqref{42w} and \eqref{41w} can be rewritten as
\begin{widetext}
\begin{align}
&\left[\phi_Xe^{\nu-\mu_2}\left(re^\nu B\right)_{,r}\right]_{,r}+\left[\omega^2\phi_Xre^{\mu_2}-\left(\mu^2+2\right)\phi_X\frac{e^{2\nu+\mu_2}}{r}-\frac{4Q_*^2}{f_Rr^3}e^{2\nu+\mu_2}\right]B=\frac{e^{\nu+\mu_2}Q_*}{f_Rr^4}Q\,,\label{471}\\
&\left(\frac{e^{\nu-\mu_2}}{f_Rr^2}Q_{,r}\right)_{,r}+\left(\frac{\omega^2}{f_Rr^2e^{\nu-\mu_2}}-\frac{e^{\nu+\mu_2}\mu^2}{f_Rr^4}\right)Q=\frac{4e^{2\nu+\mu_2}Q_*\mu^2}{f_Rr^3}B\,,\label{461}
\end{align}
\end{widetext}
where we have used Eq.~\eqref{backgroundFF} for the background field $F_{(0)(2)}$ and the Fourier decomposition $\partial_t\rightarrow-i\omega$.

We introduce the following definitions
\begin{equation}
H_1^{(-)}\equiv-2\mu\phi_X^{1/2}re^\nu B\,,\qquad H_2^{(-)}\equiv \frac{Q}{Z}\,,\label{subfr1}
\end{equation}
where $Z\equiv rf_R^{1/2}$, and consider the tortoise radius $r_*$ which satisfies
\begin{equation}
\frac{dr}{dr_*}=e^{\nu-\mu_2}\,.\label{subfr2}
\end{equation}
Finally, by using Eqs.~\eqref{subfr1} and \eqref{subfr2}, Eqs.~\eqref{471} and \eqref{461} become
\begin{widetext}
\begin{align}
\frac{d^2H_1^{(-)}}{dr_*^2}+\omega^2H_1^{(-)}=&\,\left[\frac{1}{2\phi_X^{1/2}}\left(\frac{\phi_{X,r_*}}{\phi_X^{1/2}}\right)_{,r_*}+\left(\mu^2+2\right)\frac{e^{2\nu}}{r^2}+\frac{4Q_*^2}{r^4f_R\phi_X}e^{2\nu}\right]H_1^{(-)}-\frac{2\mu e^{2\nu}Q_*}{f_R^{1/2}\phi_X^{1/2}r^3}H_2^{(-)}\,,\label{fr222}\\
\frac{d^2H_2^{(-)}}{dr_*^2}+\omega^2H_2^{(-)}=&\,\left[-Z\left(\frac{Z_{,r_*}}{Z^2}\right)_{,r_*}+\frac{e^{2\nu}\mu^2}{r^2}\right]H_2^{(-)}-\frac{2\mu e^{2\nu}Q_*}{f_R^{1/2}\phi_X^{1/2}r^3}H_1^{(-)}\,.\label{fr111}
\end{align}
\end{widetext}
It can be seen that the coupled master equations have been recast into a Schr\"odinger-like form and they can be written in a matrix expression as follows
\begin{equation}
\left(\frac{d^2}{dr_*^2}+\omega^2\right)
\begin{bmatrix}
    H_1^{(-)}  \\
    H_2^{(-)}
    \end{bmatrix}=
    \begin{bmatrix}
    V_{11} &V_{12} \\
    V_{21} &V_{22}
    \end{bmatrix}
    \begin{bmatrix}
    H_1^{(-)}  \\
    H_2^{(-)}
    \end{bmatrix}\,,\label{coupledeq1}
\end{equation}
where $V_{ij}$ is given in Eqs.~\eqref{fr222} and \eqref{fr111}.

According to the coupled master equations \eqref{fr222} and \eqref{fr111}, one can see that:
\begin{enumerate}[(i)]
\item When $f_R=1$ and the NED model is assumed to be the Born-Infeld NED given by Eq.~\eqref{BINEDla}, the master equations reduce to those of the EBI black hole given in Ref.~\cite{Fernando:2004pc}.
\item When $\phi=X$, it can be proven that the Ricci scalar $R$, the function $f$, and its derivative $f_R$ are just constants at the background level. They manifest themselves as an effective cosmological constant $\Lambda_{eff}\equiv f/(2f_R)$. Furthermore, it can be shown that, after a constant rescaling of $H_i^{(-)}$, the master equations reduce to those of the RN-dS(AdS) spacetime given in Refs.~\cite{Wang:2000gsa,Wang:2000dt,Mellor:1989ac,Molina:2003ff,Jing:2003wq}.
\item If $\phi=X$, and $f_R=1$, we have 
\begin{align}
V_{12}&=V_{21}=-\frac{2Q_*\mu}{r^3}e^{2\nu}\,,\\
V_{11}&=\frac{e^{2\nu}}{r^3}\left[(\mu^2+2)r+\frac{4Q_*^2}{r}\right]\,,\\
V_{22}&=\frac{e^{2\nu}}{r^3}\left[(\mu^2+2)r-3+\frac{4Q_*^2}{r}\right]\,.
\end{align}
The master equations turn out to be those of the RN black hole \cite{Chandrabook}.
\item If $\phi=X$, $Q_*=0$, and $f_R=1$, we have $V_{12}=V_{21}=0$, and
\begin{align}
V_{11}&=\frac{e^{2\nu}}{r^2}l(l+1)\,,\label{v11sw}\\
V_{22}&=\frac{e^{2\nu}}{r^2}\left[l(l+1)-\frac{3}{r}\right]\,.\label{v22sw}
\end{align}
Therefore, the potential for pure electromagnetic perturbations and for pure axial gravitational perturbations of the Schwarzschild black hole (the Regge-Wheeler equation \cite{Regge:1957td}) are recovered, respectively.
\end{enumerate}

As we show in the appendix~\ref{deapp}, the coupled master equation~\eqref{coupledeq1} can be decoupled within a WKB approximation in the cases that we are studying in this work. Therefore, to proceed we will decouple the master equation by diagonalizing the matrix $V_{ij}$ to obtain its eigenvalues $V_1$ and $V_2$: 
\begin{align}
V_1&=\frac{1}{2}\left[V_{11}+V_{22}+\sqrt{\left(V_{11}-V_{22}\right)^2+4V_{12}V_{21}}\right]\,,\nonumber\\
V_2&=\frac{1}{2}\left[V_{11}+V_{22}-\sqrt{\left(V_{11}-V_{22}\right)^2+4V_{12}V_{21}}\right]\,.\nonumber\\\label{diagonizingmatrix}
\end{align}

If $\phi=X$, $Q_*=0$, and $f_R=1$, it can be seen that $V_1$ and $V_2$ reduce to Eqs.~\eqref{v11sw} and \eqref{v22sw}, respectively. From the discussion above, we have proven that the potential terms in the master equation \eqref{coupledeq1} reduce to their RN counterpart in the correct limits. In the presence of nonlinearity of NED and the gravitational function $f(R)$, the potentials would change significantly and, consequently, alter the QNM frequencies. We will discuss this issue later in section \ref{sectV}.

\subsection{Axial perturbations for EiBI charged black holes}
In this subsection, we will consider the EiBI theory coupled with linear Maxwell fields. The total action is given by \cite{Banados:2010ix}
\begin{equation}
\mathcal{S}_2=\frac{\epsilon\beta_g^2}{8\pi}\int d^4x\left(\sqrt{\left|g_{\mu\nu}+\frac{R_{(\mu\nu)}}{\epsilon\beta_g^2}\right|}-\lambda\sqrt{-g}\right)+\mathcal{S}_m\,,
\end{equation}
where the matter Lagrangian is described by the linear Maxwell fields: $\phi_X=1$. In the above action, $\epsilon=\pm1$ indicates that one can freely choose the Born-Infeld coupling constant to be either positive or negative. The dimensionless constant $\lambda$ is related to an effective cosmological constant via $\Lambda=\epsilon\beta_g^2(\lambda-1)$. In the rest of this paper, we will assume a zero effective cosmological constant ($\lambda=1$) and focus on black hole solutions which are asymptotically flat. It should be stressed that only the symmetric part of the Ricci tensor $R_{(\mu\nu)}(\Gamma)$ is considered to respect the projective symmetry of the theory.

Since only the linear Maxwell fields are considered, the perturbed Maxwell equation can be obtained by simply rewriting the NED equation \eqref{21} with $\phi=X$:
\begin{align}
&\left[e^{\nu-\mu_2}(re^\nu B)_{,r}\right]_{,r}+\frac{e^{2\nu+\mu_2}}{r}\left(\frac{B_{,\theta}}{\sin{\theta}}\right)_{,\theta}\sin{\theta}-re^{\mu_2}B_{,00}\nonumber\\=&-Q_*\left(\sigma_{,20}-q_{2,00}\right)\sin^2{\theta}\,.\label{linearcharge}
\end{align}
On the other hand, one needs to take the perturbed gravitational equation into account to complete the derivation of the coupled master equations of the EiBI theory. The perturbed gravitational equation contains the perturbed energy momentum tensor, which can be rewritten from that of the NED with $\phi=X$ as:
\begin{align}
\delta T_{(0)(0)}&=-\delta T_{(2)(2)}=\frac{1}{4\pi}F_{(0)(2)}\delta F_{(0)(2)}\,,\\
\delta T_{(3)(3)}&=\delta T_{(1)(1)}=\frac{1}{4\pi}F_{(0)(2)}\delta F_{(0)(2)}\,,\\
\delta T_{(0)(1)}&=\frac{1}{4\pi}F_{(0)(2)}F_{(1)(2)}\,,\nonumber\\\delta T_{(0)(3)}&=-\frac{1}{4\pi}F_{(0)(2)}F_{(2)(3)}\,,\nonumber\\
\delta T_{(1)(2)}&=-\frac{1}{4\pi}F_{(0)(2)}F_{(0)(1)}\,,\nonumber\\\delta T_{(2)(3)}&=-\frac{1}{4\pi}F_{(0)(2)}F_{(0)(3)}\,,\\
\delta T_{(0)(2)}&=\delta T_{(1)(3)}=0\,.
\end{align}

\subsubsection{Perturbed field equations}
Because the EiBI gravity is also formulated within the Palatini variational principle, there is an auxiliary metric $q_{\mu\nu}$ which is compatible with the affine connection as before. The line element of $q_{\mu\nu}$ can be similarly expressed as in Eq.~\eqref{qmetric}. According to the variation of the action, the auxiliary metric satisfies the following equation \cite{Banados:2010ix}:
\begin{equation}
R_{(\mu\nu)}(q)=\epsilon\beta_g^2\left(q_{\mu\nu}-g_{\mu\nu}\right)\,.\label{rqgoriginal}
\end{equation}
At the beginning, the symmetric part of the Ricci tensor is assumed to be constructed from the affine connection. However, we can recast it as a function of the auxiliary metric, that is, $R_{(\mu\nu)}(q)$, because the compatibility between the affine connection and the auxiliary metric.

In the tetrad frame, we can rewrite Eq.~\eqref{rqgoriginal} as
\begin{equation}
{\tilde{e}^{(a)}}_{\mu}{\tilde{e}^{(b)}}_{\nu}\tilde{R}_{(a)(b)}=\epsilon\beta_g^2\eta_{(a)(b)}\left({\tilde{e}^{(a)}}_{\mu}{\tilde{e}^{(b)}}_{\nu}-e^{(a)}_{\mu}e^{(b)}_{\nu}\right)\,,
\end{equation}
or 
\begin{equation}
\tilde{R}_{(a)(b)}=\epsilon\beta_g^2\left(\eta_{(a)(b)}-\eta_{(c)(d)}e^{(c)}_\mu e^{(d)}_\nu{\tilde{e}^{\mu}}_{(a)}{\tilde{e}^{\nu}}_{(b)}\right)\,.\label{EiBIR}
\end{equation}
Using a similar procedure to the one we have carried previously in the Palatini $f(R)$ gravity, we have constructed another tetrad basis $\tilde{e}^{(a)}_\mu$ to map the auxiliary metric $q_{\mu\nu}$ onto the tetrad frame. 

We next consider the other field equation given in Ref.~\cite{Banados:2010ix}
\begin{equation}
\sqrt{-q}q^{\mu\nu}-\sqrt{-g}g^{\mu\nu}=-\frac{8\pi}{\epsilon\beta_g^2}\sqrt{-g}T^{\mu\nu}\,.
\end{equation}
In the tetrad frame, this equation reads
\begin{align}
&e\left(\eta_{(a)(b)}-\frac{8\pi}{\epsilon\beta_g^2}T_{(a)(b)}\right)\nonumber\\=&\,\tilde{e}\eta^{(g)(h)}\eta_{(a)(c)}\eta_{(b)(d)}{\tilde{e}^{\mu}}_{(g)}{\tilde{e}^{\nu}}_{(h)}e^{(c)}_{\mu}e^{(d)}_{\nu}\,,\label{EiBIgq}
\end{align}
where $e=\sqrt{-g}$ and $\tilde{e}=\sqrt{-q}$. Even though the two tetrad bases are not related explicitly as in the Palatini $f(R)$ gravity, that is Eq.~\eqref{eefRrelation}, in the EiBI theory these two bases are still related implicitly via Eqs.~\eqref{EiBIR} and \eqref{EiBIgq}. We will immediately show how to derive the master equations of the axial perturbations by using these two equations.

The $(1,3)$ component of Eqs.~\eqref{EiBIR} and \eqref{EiBIgq} can be explicitly written as
\begin{align}
\delta R_{(1)(3)}&=-\epsilon\beta_g^2e^{(1)}_1{\tilde{e}^{1}}_{(1)}\left(e^{(1)}_1{\tilde{e}^{1}}_{(3)}+e^{(1)}_3{\tilde{e}^{3}}_{(3)}\right)\,,\\
0&=\tilde{e}e^{(3)}_3{\tilde{e}^{3}}_{(3)}\left(e^{(1)}_1{\tilde{e}^{1}}_{(3)}+e^{(1)}_3{\tilde{e}^{3}}_{(3)}\right)\,,
\end{align}
where the 1st order terms are collected inside the parenthesis for each equation (see the components of a tetrad basis explicitly given in Eqs.~\eqref{tetradbasis111} and \eqref{tetradbasis222}). One can then obtain
\begin{align}
&\delta R_{(1)(3)}\nonumber\\
=&\,\left[e^{3\tilde{\psi}+\tilde\nu-\tilde{\mu}_3-\tilde{\mu}_2}\left(\tilde{q}_{2,3}-\tilde{q}_{3,2}\right)\right]_{,2}-e^{3\tilde{\psi}-\tilde\nu+\tilde{\mu}_2-\tilde{\mu}_3}\left(\tilde{\sigma}_{,3}-\tilde{q}_{3,0}\right)_{,0}\nonumber\\=&\,0\,.\label{71}
\end{align}

Next, the $(1,2)$ component of Eqs.~\eqref{EiBIR} and \eqref{EiBIgq} can be explicitly written as
\begin{align}
\delta R_{(1)(2)}&=-\epsilon\beta_g^2e^{(1)}_1{\tilde{e}^{1}}_{(1)}\left(e^{(1)}_1{\tilde{e}^{1}}_{(2)}+e^{(1)}_2{\tilde{e}^{2}}_{(2)}\right)\,,\\
-\frac{8\pi e}{\epsilon\beta_g^2}\delta T_{(1)(2)}&=\tilde{e}e^{(2)}_2{\tilde{e}^{2}}_{(2)}\left(e^{(1)}_1{\tilde{e}^{1}}_{(2)}+e^{(1)}_2{\tilde{e}^{2}}_{(2)}\right)\,.\label{73}
\end{align}
We then have
\begin{align}
\delta R_{(1)(2)}&=8\pi\frac{e}{\tilde{e}}\frac{e^{(1)}_1{\tilde{e}^{1}}_{(1)}}{e^{(2)}_2{\tilde{e}^{2}}_{(2)}}\delta T_{(1)(2)}=-\frac{2}{\sqrt{\sigma_+^3\sigma_-}}F_{(0)(2)}F_{(0)(1)}\,,\label{r12eibiinter}
\end{align}
where we have defined
\begin{equation}
\sigma_{\pm}\equiv1\pm\frac{Q_*^2}{\epsilon\beta_g^2r^4}\,.
\end{equation}
Eq.~\eqref{r12eibiinter} can be rewritten as follows
\begin{align}
&\left[e^{3\tilde{\psi}+\tilde\nu-\tilde{\mu}_3-\tilde{\mu}_2}\left(\tilde{q}_{2,3}-\tilde{q}_{3,2}\right)\right]_{,3}+e^{3\tilde{\psi}-\tilde\nu+\tilde{\mu}_3-\tilde{\mu}_2}\left(\tilde{\sigma}_{,2}-\tilde{q}_{2,0}\right)_{,0}\nonumber\\
=&\,-4e^\nu rQ_*B\sin{\theta}\,.\label{75}
\end{align}

Finally, the $(0,1)$ component of Eqs.~\eqref{EiBIR} and \eqref{EiBIgq} can be explicitly written as
\begin{align}
\delta R_{(0)(1)}&=-\epsilon\beta_g^2e^{(1)}_1{\tilde{e}^{1}}_{(1)}\left(e^{(1)}_1{\tilde{e}^{1}}_{(0)}+e^{(1)}_0{\tilde{e}^{0}}_{(0)}\right)\,,\\
-\frac{8\pi e}{\epsilon\beta_g^2}\delta T_{(0)(1)}&=\tilde{e}e^{(0)}_0{\tilde{e}^{0}}_{(0)}\left(e^{(1)}_1{\tilde{e}^{1}}_{(0)}+e^{(1)}_0{\tilde{e}^{0}}_{(0)}\right)\,.\label{77}
\end{align}
We then have
\begin{align}
\delta R_{(0)(1)}&=8\pi\frac{e}{\tilde{e}}\frac{e^{(1)}_1{\tilde{e}^{1}}_{(1)}}{e^{(0)}_0{\tilde{e}^{0}}_{(0)}}\delta T_{(0)(1)}=\frac{2}{\sqrt{\sigma_+^3\sigma_-}}F_{(0)(2)}F_{(1)(2)}\,,
\end{align}
which leads to
\begin{align}
&\left[e^{3\tilde{\psi}-\tilde\nu-\tilde{\mu}_2+\tilde{\mu}_3}\left(\tilde{\sigma}_{,2}-\tilde{q}_{2,0}\right)\right]_{,2}+\left[e^{3\tilde{\psi}-\tilde\nu-\tilde{\mu}_3+\tilde{\mu}_2}\left(\tilde{\sigma}_{,3}-\tilde{q}_{3,0}\right)\right]_{,3}\nonumber\\=&\,4e^{\mu_2} rQ_*\sin^2{\theta}F_{(1)(2)}\,.
\end{align}

To simplify the equations, we define 
\begin{align}
\tilde{Q}&\equiv e^{3\tilde{\psi}+\tilde\nu-\tilde{\mu}_3-\tilde{\mu}_2}\left(\tilde{q}_{2,3}-\tilde{q}_{3,2}\right)\nonumber\\&=r^2e^{\nu-\mu_2}\sigma_+\sin^3{\theta}\left(\tilde{q}_{2,3}-\tilde{q}_{3,2}\right)\,.\label{definQeibi}
\end{align}
Then, Eqs.~\eqref{71} and \eqref{75} become
\begin{align}
&e^{\nu-\mu_2}\frac{\tilde{Q}_{,2}}{\sigma_+r^2\sin^3{\theta}}=\left(\tilde{\sigma}_{,3}-\tilde{q}_{3,0}\right)_{,0}\,,\label{82previous}\\
&e^{\nu+\mu_2}\left(\frac{\sigma_-}{\sigma_+^2}\right)\frac{\tilde{Q}_{,3}}{r^4\sin^3{\theta}}\nonumber\\=&-(\tilde{\sigma}_{,2}-\tilde{q}_{2,0})_{,0}-\frac{4Q_*}{r^3\sin^2{\theta}}\left(\frac{\sigma_-}{\sigma_+^2}\right)e^{2\nu+\mu_2}B\,.\label{82}
\end{align}
By differentiating Eqs.~\eqref{82previous} and \eqref{82} and eliminating $\tilde\sigma$, we have
\begin{align}
&\frac{1}{\sin^3{\theta}}\left(\frac{e^{\nu-\mu_2}}{\sigma_+r^2}\tilde{Q}_{,2}\right)_{,2}+\frac{e^{\nu+\mu_2}}{r^4}\left(\frac{\sigma_-}{\sigma_+^2}\right)\left(\frac{\tilde{Q}_{,3}}{\sin^3{\theta}}\right)_{,3}\nonumber\\
=&\,\frac{\tilde{Q}_{,00}}{\sigma_+r^2\sin^3{\theta}e^{\nu-\mu_2}}-\frac{4e^{2\nu+\mu_2}Q_*}{r^3}\left(\frac{\sigma_-}{\sigma_+^2}\right)\left(\frac{B}{\sin^2\theta}\right)_{,3}\,.\label{83w}
\end{align}
We have obtained one of the coupled master equations \eqref{83w}. Now we need to consider the other equation which comes from the perturbed Maxwell equation, that is, Eq.~\eqref{linearcharge}. It is necessary to replace the right hand side of Eq.~\eqref{linearcharge} with its tilde counterpart because of the definition \eqref{definQeibi}.

If we express Eqs.~\eqref{73} and \eqref{77} more explicitly by writing down the metric functions, we obtain
\begin{align}
q_2&=\tilde{q}_2-\frac{2e^{\mu_2}Q_*}{\epsilon\beta_g^2r^3\sigma_+}\frac{F_{(0)(1)}}{\sin{\theta}}\,,\\
\sigma&=\tilde{\sigma}+\frac{2e^{\nu}Q_*}{\epsilon\beta_g^2r^3\sigma_+}\frac{F_{(1)(2)}}{\sin{\theta}}\,,
\end{align}
respectively. Considering the difference between $\sigma_{,20}-q_{2,00}$ and $\tilde{\sigma}_{,20}-\tilde{q}_{2,00}$, we have
\begin{align}
&\sigma_{,20}-q_{2,00}\nonumber\\=&\,\tilde{\sigma}_{,20}-\tilde{q}_{2,00}\nonumber\\
&-\frac{2Q_*}{\epsilon\beta_g^2\sin^2\theta}\left\{\left[\frac{e^{\nu-\mu_2}}{r^4\sigma_+}(re^\nu B)_{,r}\right]_{,r}-\frac{e^{\mu_2}}{r^3\sigma_+}B_{,00}\right\}\,,\label{86}
\end{align}
where we have used Eq.~\eqref{12} to replace $F_{(1)(2),0}$ with $F_{(0)(1),2}$. After combining Eqs.~\eqref{linearcharge}, \eqref{82} and \eqref{86}, we obtain
\begin{align}
&\left[e^{\nu-\mu_2}\left(\frac{\sigma_-}{\sigma_+}\right)(re^\nu B)_{,r}\right]_{,r}+\frac{e^{2\nu+\mu_2}}{r}\left(\frac{B_{,\theta}}{\sin\theta}\right)_{,\theta}\sin\theta\nonumber\\&+\left(\frac{\sigma_-}{\sigma_+}\right)\left(\omega^2re^{\mu_2}-\frac{4Q_*^2}{r^3\sigma_+}e^{2\nu+\mu_2}\right)B\nonumber\\
=&\,Q_*e^{\nu+\mu_2}\left(\frac{\sigma_-}{\sigma_+^2}\right)\frac{\tilde{Q}_{,\theta}}{r^4\sin\theta}\,.\label{87}
\end{align}

\subsubsection{Effective potentials}
Similar to what we have done in the previous subsection, we substitute \eqref{43} into the master equations. Consequently, Eqs.~\eqref{83w} and \eqref{87} can be rewritten as follows
\begin{widetext}
\begin{align}
&\left(\frac{e^{\nu-\mu_2}}{\sigma_+r^2}\tilde{Q}_{,r}\right)_{,r}+\left[\frac{\omega^2e^{-\nu+\mu_2}}{\sigma_+r^2}-\frac{\mu^2e^{\nu+\mu_2}}{r^4}\left(\frac{\sigma_-}{\sigma_+^2}\right)\right]\tilde{Q}=\frac{4\mu^2e^{2\nu+\mu_2}Q_*}{r^3}\left(\frac{\sigma_-}{\sigma_+^2}\right)B\,,\\
&\left[e^{\nu-\mu_2}\left(\frac{\sigma_-}{\sigma_+}\right)(re^\nu B)_{,r}\right]_{,r}+\left[\left(\frac{\sigma_-}{\sigma_+}\right)\left(\omega^2re^{\mu_2}-\frac{4Q_*^2}{r^3\sigma_+}e^{2\nu+\mu_2}\right)-(\mu^2+2)\frac{e^{2\nu+\mu_2}}{r}\right]B
=Q_*\left(\frac{\sigma_-}{\sigma_+^2}\right)e^{\nu+\mu_2}\frac{\tilde{Q}}{r^4}\,.
\end{align}
\end{widetext}
With the further definitions
\begin{equation}
H_1^{(-)}\equiv-2\mu\mathcal{S}re^\nu B\,,\qquad H_2^{(-)}\equiv\frac{\tilde{Q}}{W}\,,
\end{equation}
where $W\equiv r\sqrt{\sigma_+}$ and $\mathcal{S}\equiv\sqrt{\sigma_-/\sigma_+}$, and after introducing the tortoise radius
\begin{equation}
\frac{dr}{dr_*}=e^{\nu-\mu_2}\,,
\end{equation}
we can obtain
\begin{widetext}
\begin{align}
\frac{d^2H_1^{(-)}}{dr_*^2}+\omega^2H_1^{(-)}=&\,\left[\frac{\mathcal{S}_{r_*r_*}}{\mathcal{S}}+(\mu^2+2)\frac{e^{2\nu}}{r^2}\left(\frac{\sigma_+}{\sigma_-}\right)+\frac{4Q_*^2}{r^4\sigma_+}e^{2\nu}\right]H_1^{(-)}-\frac{2Q_*\mu e^{2\nu}\sqrt{\sigma_-}}{\sigma_+r^3}H_2^{(-)}\,,\label{H1EIBIEQ}\\
\frac{d^2H_2^{(-)}}{dr_*^2}+\omega^2H_2^{(-)}=&\,\left[-W\left(\frac{W_{,r_*}}{W^2}\right)_{,r_*}+\frac{e^{2\nu}\mu^2}{r^2}\left(\frac{\sigma_-}{\sigma_+}\right)\right]H_2^{(-)}-\frac{2Q_*\mu e^{2\nu}\sqrt{\sigma_-}}{\sigma_+r^3}H_1^{(-)}\,.\label{H2EIBIEQ}
\end{align}
\end{widetext}
These coupled equations can be written in a matrix form similar to the one given in Eq.~\eqref{coupledeq1}. Furthermore, since the eigenvectors of the matrix $V_{ij}$ are approximately constant as those for the EBI black hole shown in the appendix~\ref{deapp}, we can diagonalize the matrix as we did in Eq.~\eqref{diagonizingmatrix}. Because the expressions are so complicated, we did not write down the explicit form of $V_i$ in this paper. It is still important to check whether the potentials reduce to those of the RN black hole in the proper limits. Specifically, if $\beta_g^2\rightarrow\infty$ we have $\sigma_+\sim\sigma_-\sim1$ and the master equations reduce to those of the RN black hole. Moreover, if $Q_*=0$ the master equations of the Schwarzschild black hole are recovered.

In addition, we should highlight a crucial result following from the master equations \eqref{H1EIBIEQ} and \eqref{H2EIBIEQ}. It can be seen that in the second term on the right hand side of Eq.~\eqref{H1EIBIEQ}, i.e., the term containing $\mu^2$, there is a factor $\sigma_+/\sigma_-$. On the other hand, in the second term on the right hand side of Eq.~\eqref{H2EIBIEQ}, there is a factor $\sigma_-/\sigma_+$. Actually, these factors play a crucial role when the QNMs in the eikonal limit ($l\rightarrow\infty$) are considered. In that limit, we will show later in section~\ref{sectV} that, because of these factors, the QNMs cannot be calculated directly from the associated quantities of the unstable photon sphere of the black hole and the correspondence proposed in Ref.~\cite{Cardoso:2008bp} is not satisfied for the EiBI charged black holes (for more fundamental illustration on the photon sphere, see Ref.~\cite{Claudel:2000yi}).

Before closing this subsection, we would like to write down the exact metric functions of the charged black holes in the EiBI gravity at the background level:
\begin{enumerate}[(i)]
\item If $\epsilon=+1$, the metric functions read \cite{Wei:2014dka,Sotani:2014lua,Chen:2018mkf}
\begin{widetext}
\begin{align}
e^{2(\nu+\mu_2)}&=\frac{r^4}{r^4+r_g^4}\,,\nonumber\\
e^{-2\mu_2}&=\frac{r^4+r_g^4}{r^4-r_g^4}\left[1-\frac{r}{\sqrt{r^4+r_g^4}}\left(1-\frac{4r_g^4\beta_g^2}{3r}F\left(\frac{1}{4},\frac{1}{2};\frac{5}{4};-\frac{r_g^4}{r^4}\right)\right)-\frac{r_g^4\beta_g^2}{3r^2}\right]\,,\nonumber\\
e^{2\mu_3}&=r^2\,,\qquad e^{2\psi}=r^2\sin^2\theta\,,\label{EiBIplus}
\end{align}
\end{widetext}
where $r_g\equiv\sqrt{Q_*/\beta_g}$ and we have used the rescaling $\beta_gr_s\rightarrow\beta_g$.

\item If $\epsilon=-1$, the metric functions read \cite{Wei:2014dka,Sotani:2014lua,Chen:2018mkf}
\begin{widetext}
\begin{align}
e^{2(\nu+\mu_2)}&=\frac{r^4}{r^4-r_g^4}\,,\nonumber\\
e^{-2\mu_2}&=\frac{r^4-r_g^4}{r^4+r_g^4}\left[1-\frac{r}{\sqrt{r^4-r_g^4}}\left(1-\frac{r_g^3\beta_g^2}{3}B\left(\frac{1}{4},\frac{1}{2}\right)+\frac{2\sqrt{2}r_g^3\beta_g^2}{3}F\left(\cos^{-1}\frac{r_g}{r},\frac{1}{\sqrt{2}}\right)\right)-\frac{r_g^4\beta_g^2}{3r^2}\right]\,,\nonumber\\
e^{2\mu_3}&=r^2\,,\qquad e^{2\psi}=r^2\sin^2\theta\,,\label{EiBIminus}
\end{align}
\end{widetext}
where $B(..,..)$ is the Beta function and $F(..,..)$ is the elliptic function of the first kind, respectively \cite{Abramow}.
\end{enumerate}
The derivation of the exact metric functions given in Eqs.~\eqref{EiBIplus} and \eqref{EiBIminus} were first obtained in Refs.~\cite{Wei:2014dka,Sotani:2014lua}. One can also refer to Ref.~\cite{Chen:2018mkf} in which we recast the metric functions in a simpler form for calculating the QNM frequencies. Note that in the EiBI gravity, there are some regions of the parameter space where no black hole solution exists \cite{Wei:2014dka,Sotani:2014lua}. In this paper, we will only focus on the cases where the black holes exist and calculate their QNM frequencies.

\section{QNM frequencies: the 6th order WKB method}\label{sectV}
The evaluation of the QNM frequencies is essentially based on treating the master equations of the perturbations as an eigenvalue problem with proper boundary conditions. In the literature, there have been several methods to calculate the QNMs, ranging from numerical approaches \cite{Leaver:1986gd,Jansen:2017oag} to semi-analytic methods (see Refs.~\cite{Nollert:1999ji,Berti:2009kk,Konoplya:2011qq,Berti:2015itd} and references therein). In this paper, we will use a semi-analytical approach firstly formulated in the seminal paper \cite{Schutz:1985zz}. This approach is based on the WKB approximation and the QNMs can be calculated by just using a simple formula once the potential terms in the master equations are given. In Refs.~\cite{Iyer:1986np,Konoplya:2003ii}, the 1st order WKB method was extended to the 3rd and 6th orders WKB approximation, respectively. Recently, a further extension of the WKB method up to the 13th order has been proposed with the help of the Pad\'e transforms \cite{Matyjasek:2017psv}. The WKB method is expected to be accurate as long as the multipole number $l$ is larger than the overtone $n$ \cite{Berti:2009kk}. On the other hand, for astrophysical black holes, the fundamental mode $n=0$ has the longest decay time and therefore dominates the late time signal of the ringdown stage. At this regard, we will mainly focus on the fundamental mode.

The formulation of the WKB method to calculate the QNMs is essentially based on the fact that the master equations can be written like a Schr\"odinger wave equation in quantum mechanics. The potential term, in most cases (including ours), has a finite value when $r_*\rightarrow\infty$ (spatial infinity) and $r_*\rightarrow-\infty$ (at the event horizon). Furthermore, the potential has a peak at some finite $r_*$. One can then treat the problem as a quantum scattering process through a potential barrier after suitable boundary conditions for the problems are imposed. At spatial infinity, only outgoing waves moving away from the black hole exist. On the other hand, there can only exist ingoing waves moving toward the black hole at the event horizon.

The idea of the WKB method to encompass the aforementioned boundary conditions is to consider a quantum scattering process without incident waves, while the reflected and the transmitted waves have comparable amounts of amplitudes. The peak value of the effective potential $V_{\textrm{eff}}(r_*)\equiv -\omega^2+V$ is required to be slightly larger than zero in the sense that there are two classical turning points near the peak. The solutions far away from the turning points ($r_*\rightarrow\pm\infty$) are solved by using the WKB approximation up to a desired order and the boundary conditions should be taken into account. At the vicinity of the peak, the potential is expanded into a Taylor series up to a given order, and one uses a series expansion method to solve the differential equation. Finally, the numerical values of the QNM frequencies $\omega$ can be obtained by matching the solution near the peak with the solutions deduced from the WKB approximation simultaneously at the two turning points. 

In the 6th order WKB method, the WKB formula for calculating QNMs is \cite{Schutz:1985zz,Iyer:1986np,Konoplya:2003ii}
\begin{equation}
\frac{i\left(\omega^2-V_m\right)}{\sqrt{-2V_m''}}-\sum_{i=2}^6\Lambda_i=n+\frac{1}{2}\,,
\end{equation}
where the index $m$ denotes the quantities evaluated at the peak of the potential. $V_m''$ is the second order derivative of the potential with respect to $r_*$, calculated at the peak. $\Lambda_i$ are constant coefficients resulting from higher order WKB corrections. These coefficients contain the value and derivatives (up to the 12th order) of the potential at the peak.{\footnote{The explicit expressions of $\Lambda_i$ are given in Refs.~\cite{Iyer:1986np,Konoplya:2003ii} (see Eqs.~(1.5a) and (1.5b) in Ref.~\cite{Iyer:1986np}, and the appendix in Ref.~\cite{Konoplya:2003ii}).}}

\subsection{Fundamental QNMs}

\begin{figure*}[t]
\centering
\graphicspath{{fig/}}
\includegraphics[scale=0.55]{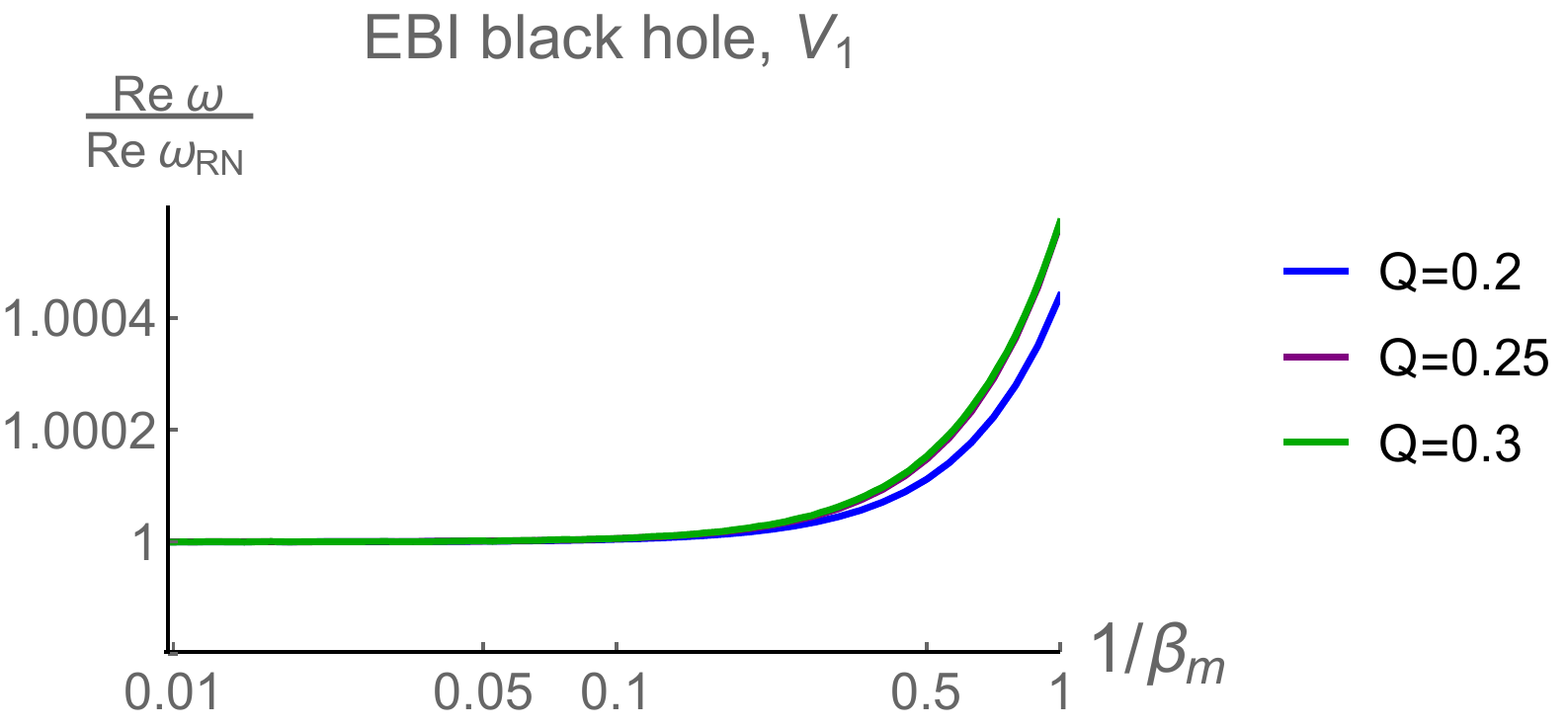}
\includegraphics[scale=0.55]{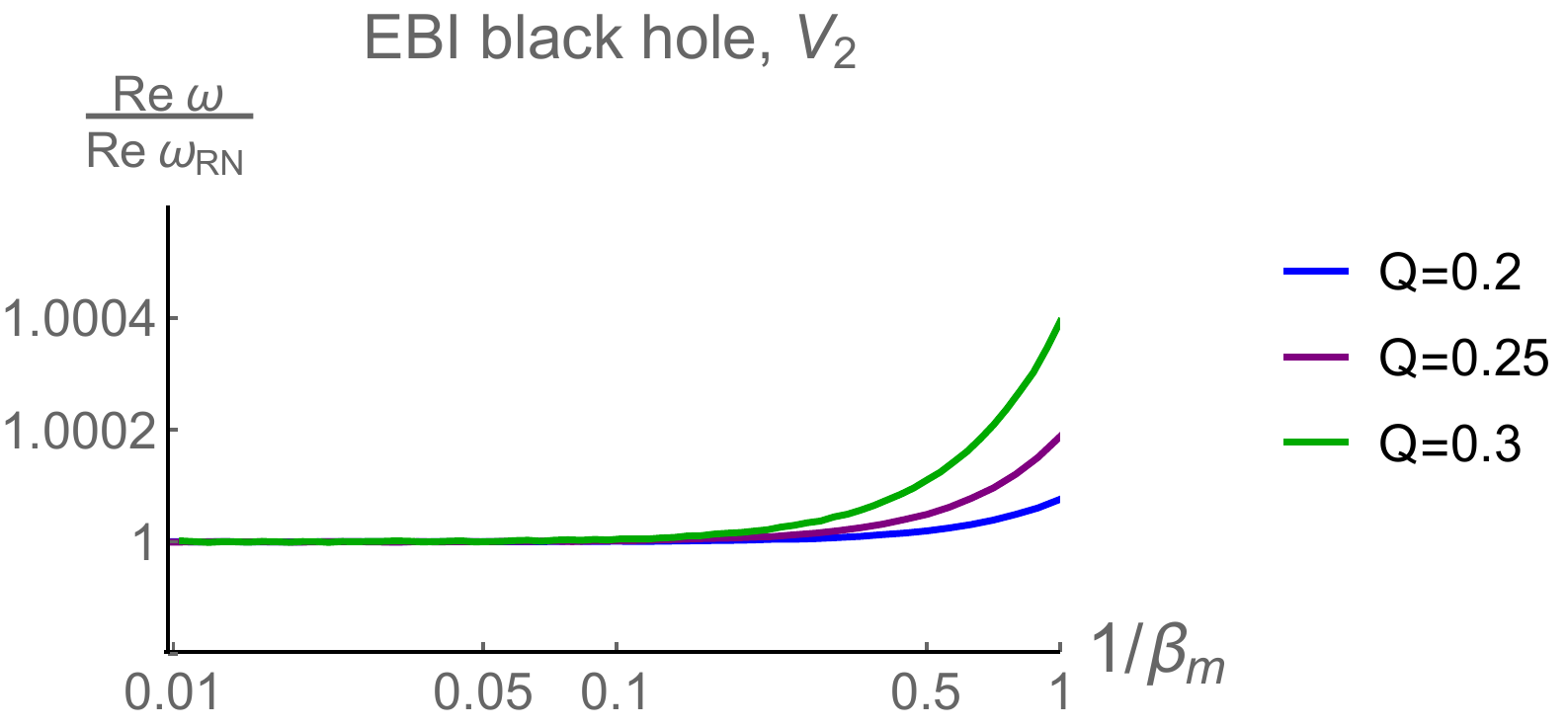}
\includegraphics[scale=0.55]{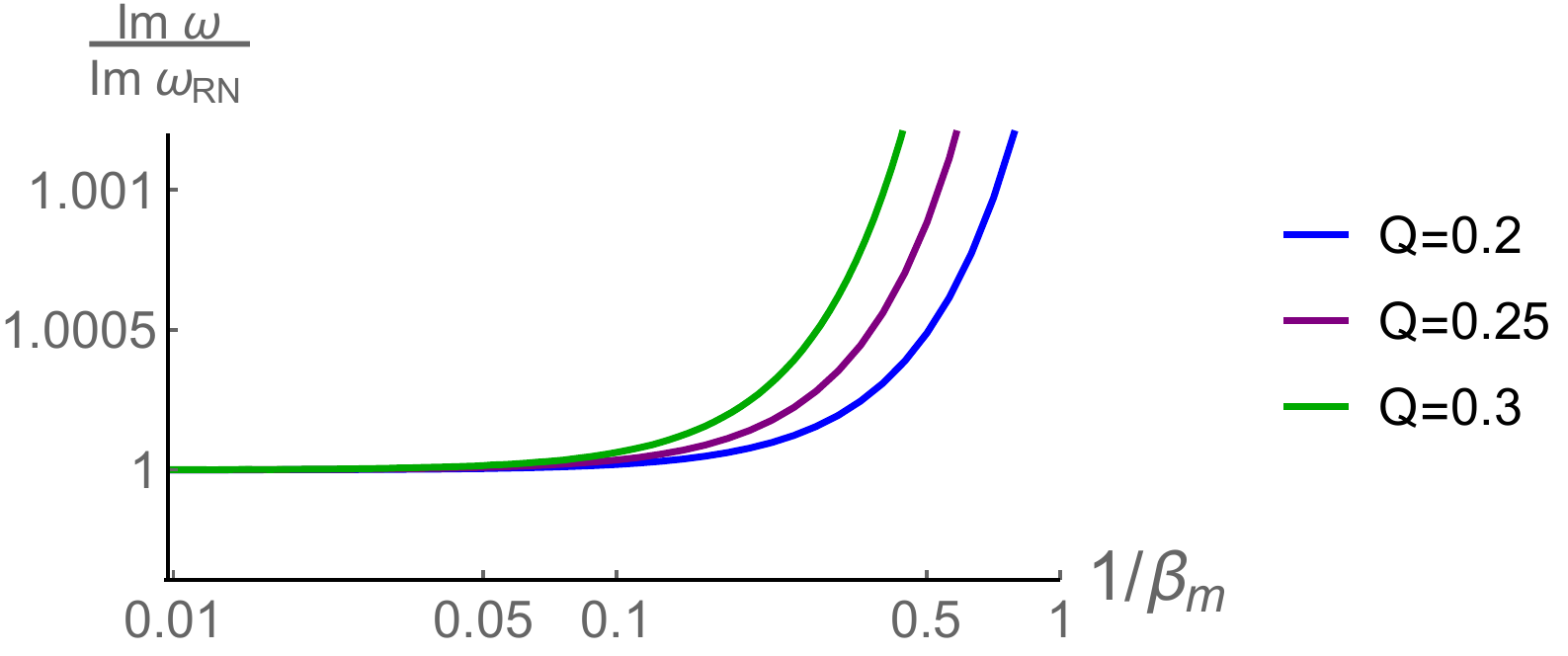}
\includegraphics[scale=0.55]{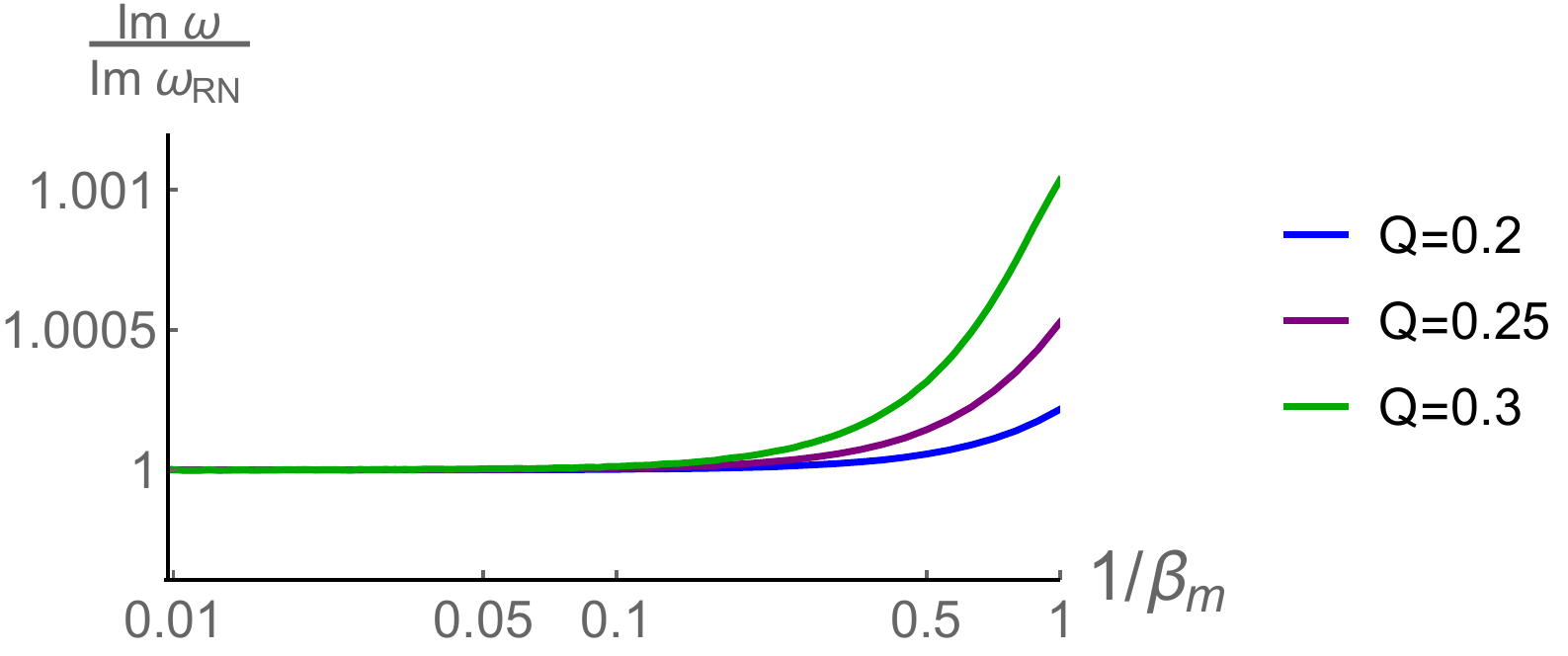}
\caption{The real part (upper) and imaginary part (lower) of the fundamental QNM frequencies of the EBI black holes are presented with respect to $1/\beta_m$. The results are based on the potential $V_1$ (left) and $V_2$ (right), and the multipole number is fixed to $l=2$.} 
\label{figEBI1}
\end{figure*}

In Fig.~\ref{figEBI1}, we consider the EBI black hole and show its QNM frequencies calculated from $V_1$ and $V_2$, respectively (see the expressions of these potentials in Eq.~\eqref{diagonizingmatrix}). We consider the multiple $l=2$ and the fundamental mode $n=0$. For the sake of convenience to highlight the deviations due to the Born-Infeld corrections, we present the QNMs ratio of the EBI black hole and the RN black hole. 

\begin{figure*}[t]
\centering
\graphicspath{{fig/}}
\includegraphics[scale=0.48]{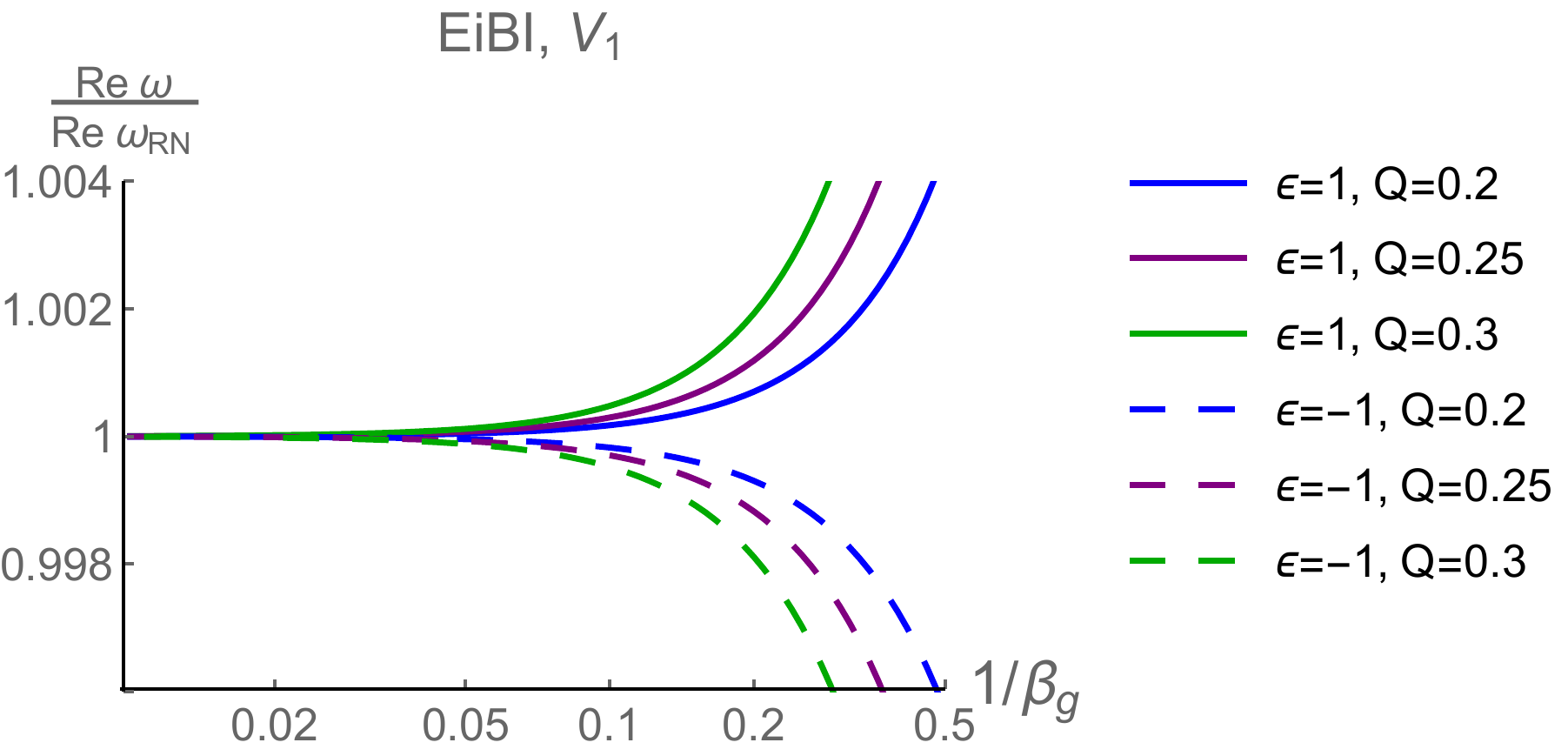}
\includegraphics[scale=0.48]{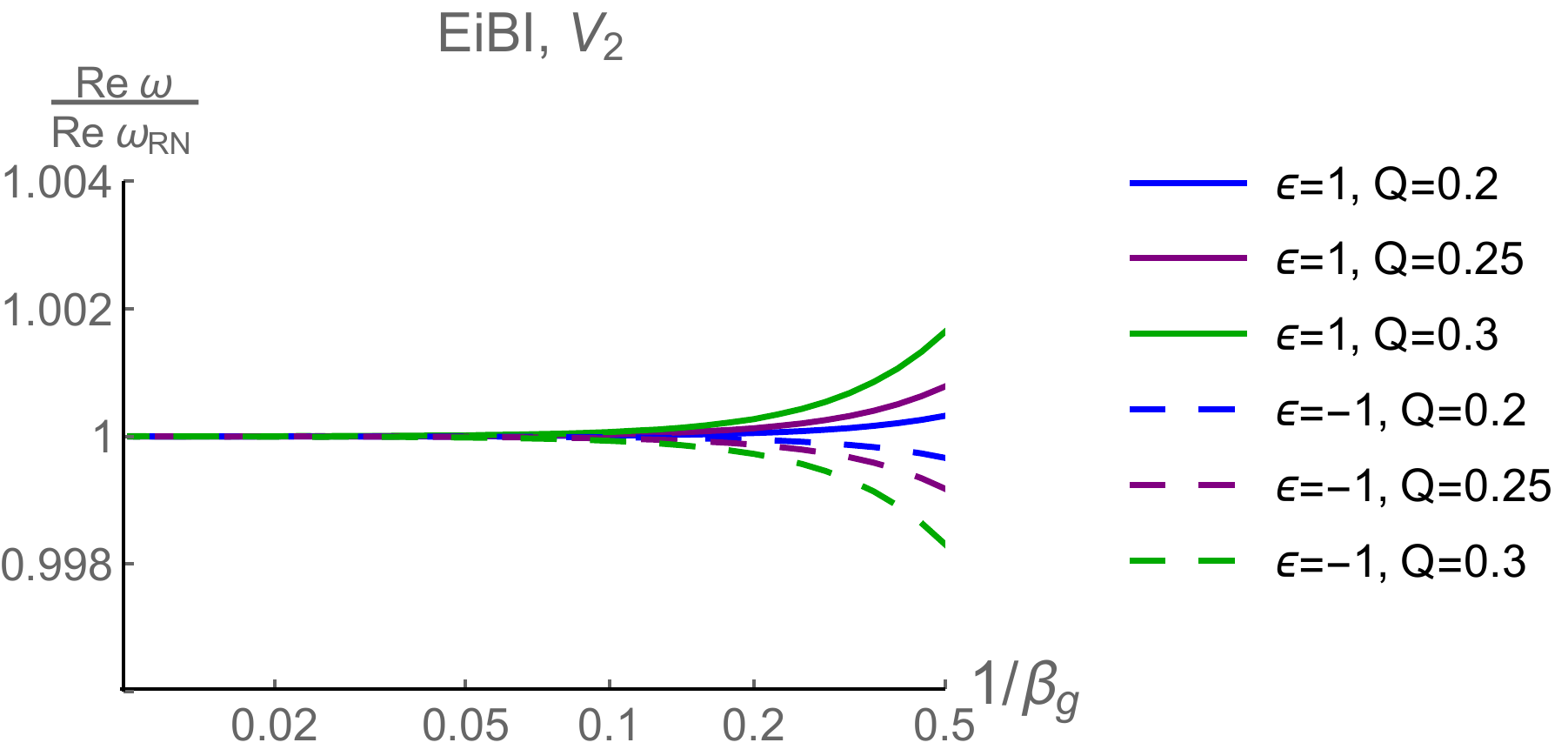}
\includegraphics[scale=0.48]{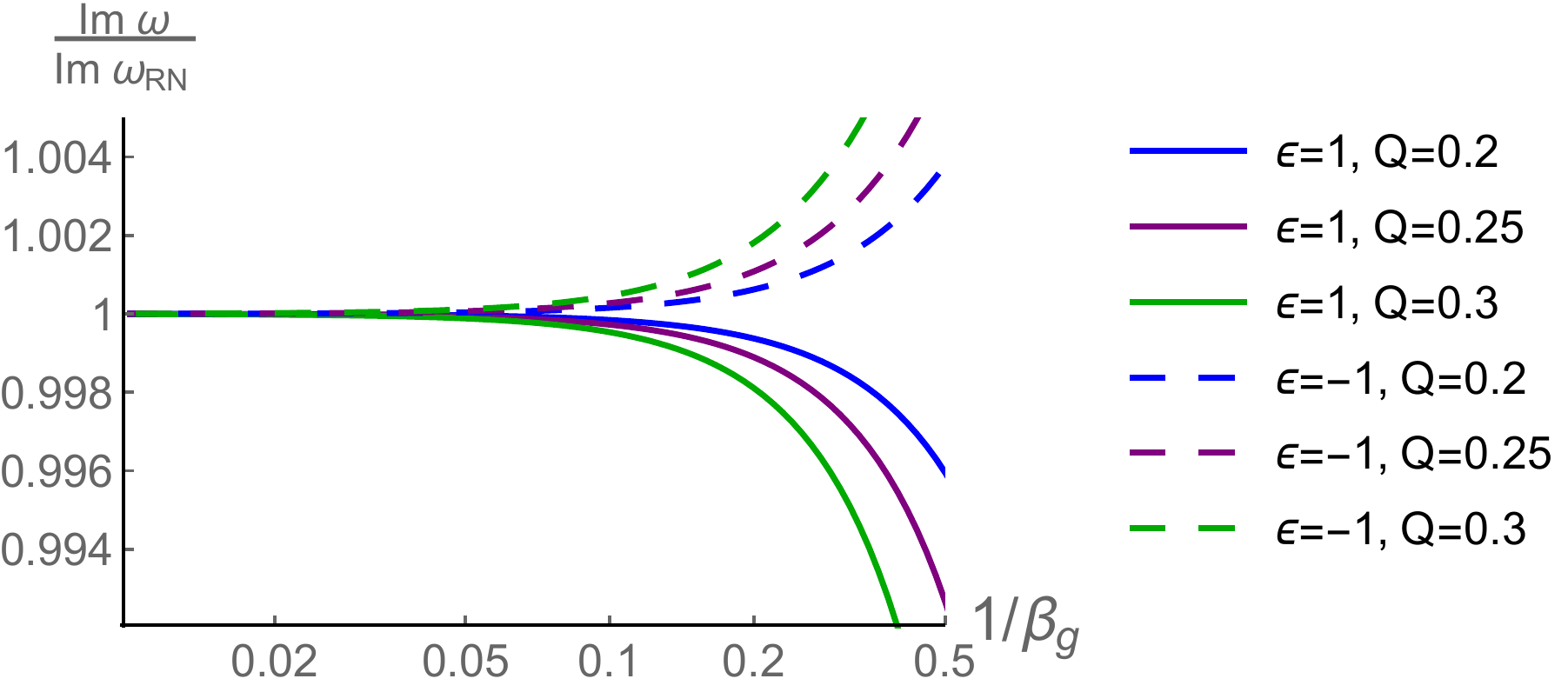}
\includegraphics[scale=0.48]{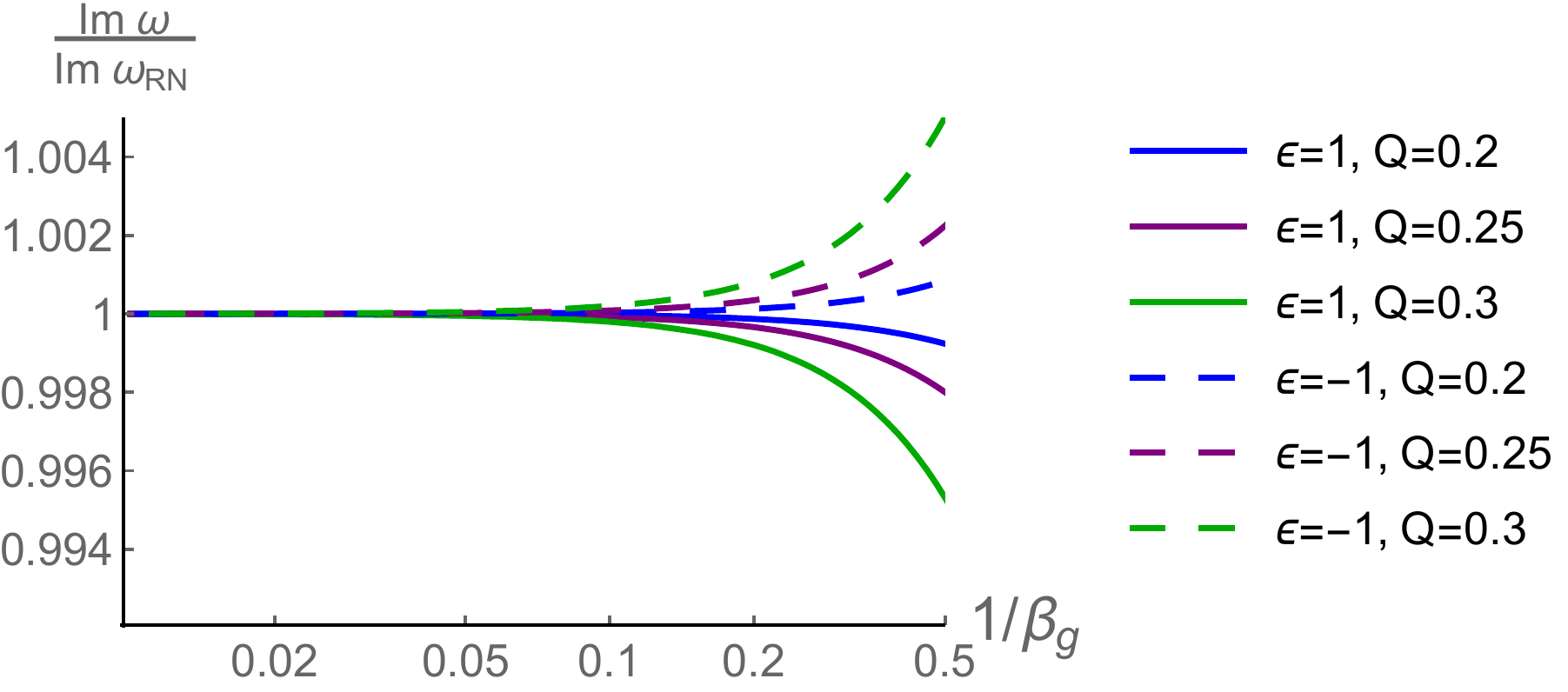}
\caption{The real part (upper) and imaginary part (lower) of the fundamental QNM frequencies of the EiBI charged black holes are presented with respect to $1/\beta_g$. The results are based on the potential $V_1$ (left) and $V_2$ (right), and the multipole number is fixed to $l=2$.} 
\label{fig1}
\end{figure*}

In Fig.~\ref{fig1}, we consider the EiBI charged black holes and exhibit their QNM frequencies calculated from $V_1$ and $V_2$, respectively. The solid curves and the dashed curves correspond to the results when the EiBI coupling constant is positive ($\epsilon=+1$) and negative ($\epsilon=-1$), respectively.

\begin{figure*}[t]
\centering
\graphicspath{{fig/}}
\includegraphics[scale=0.55]{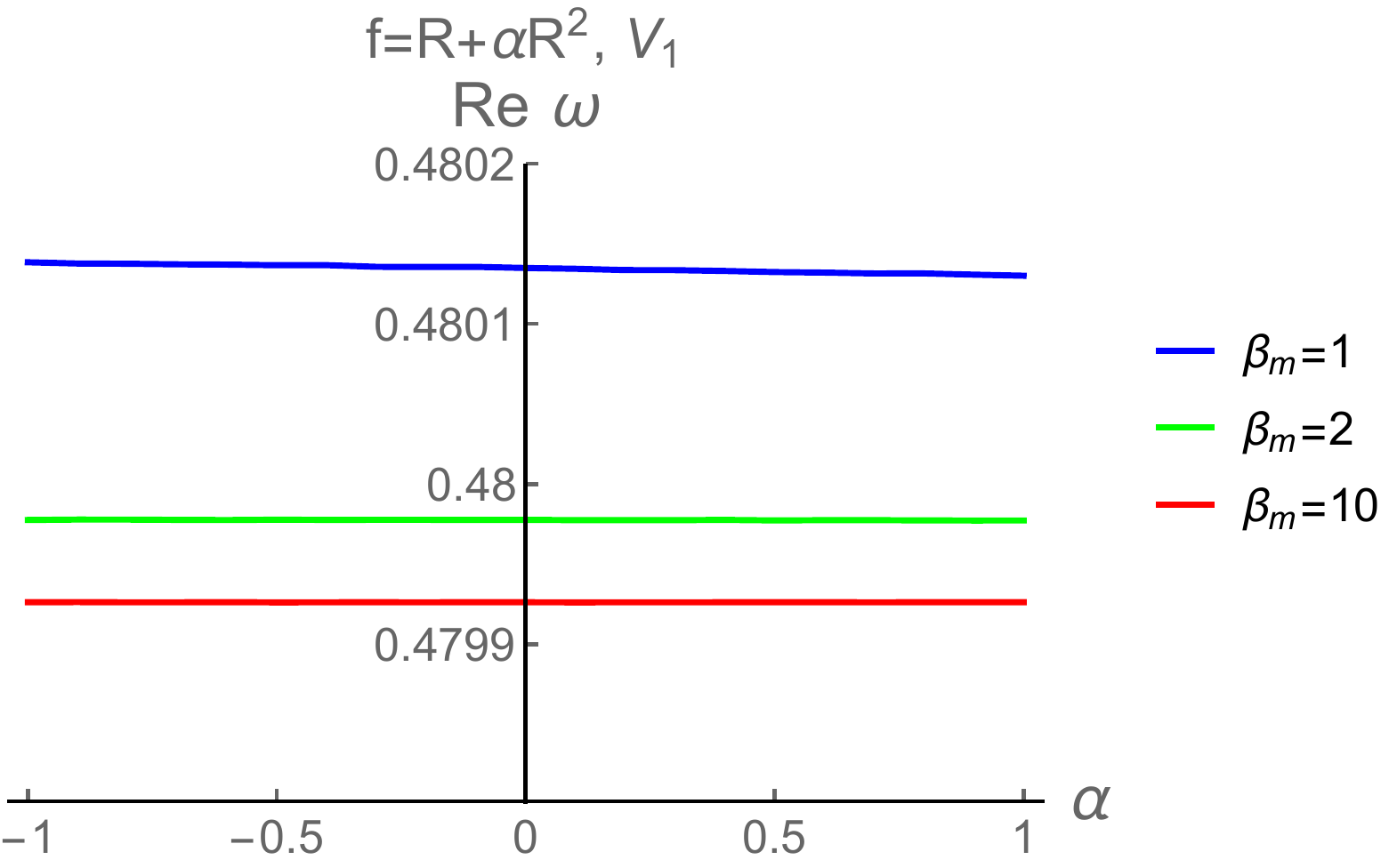}
\includegraphics[scale=0.55]{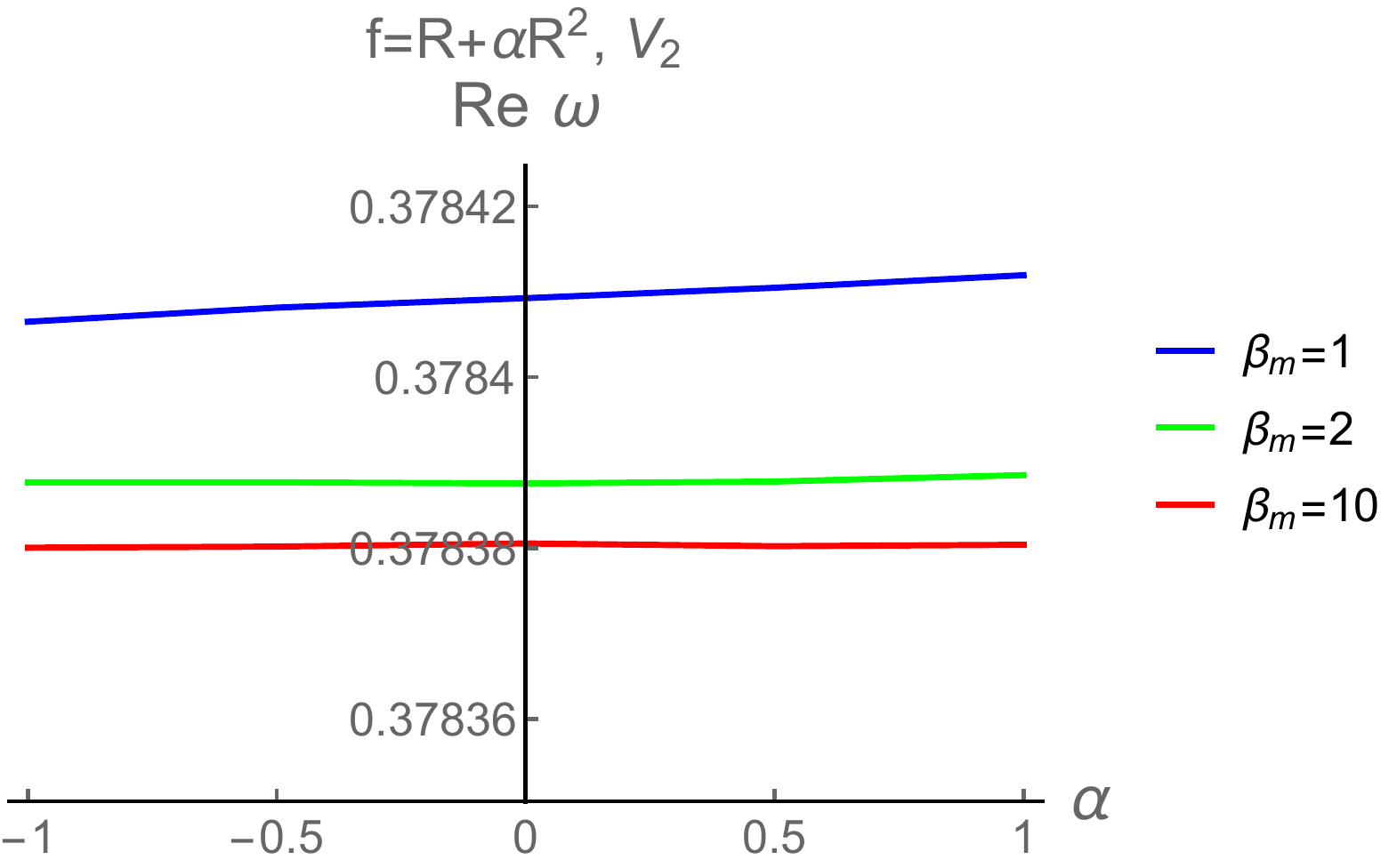}
\includegraphics[scale=0.55]{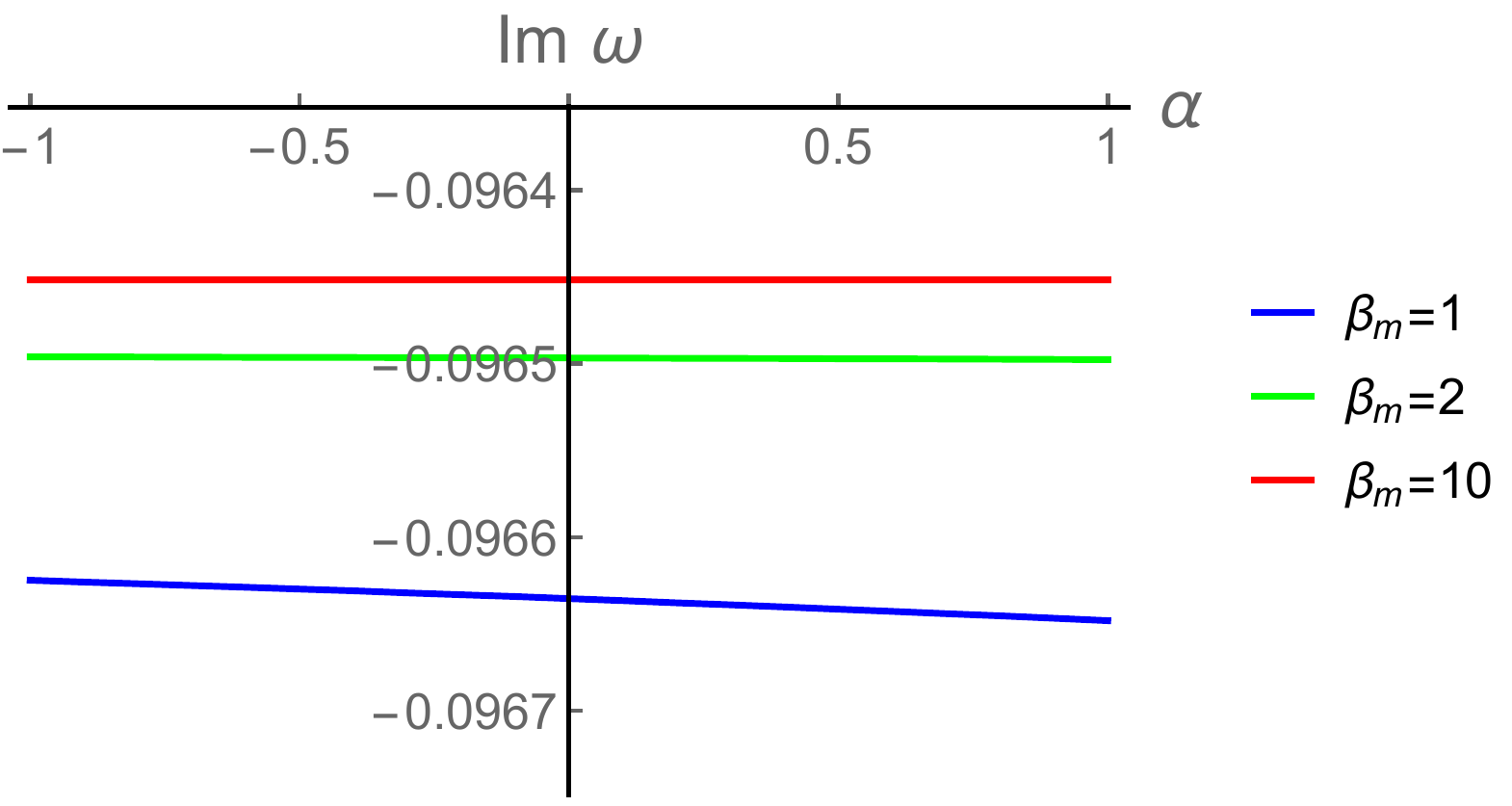}
\includegraphics[scale=0.55]{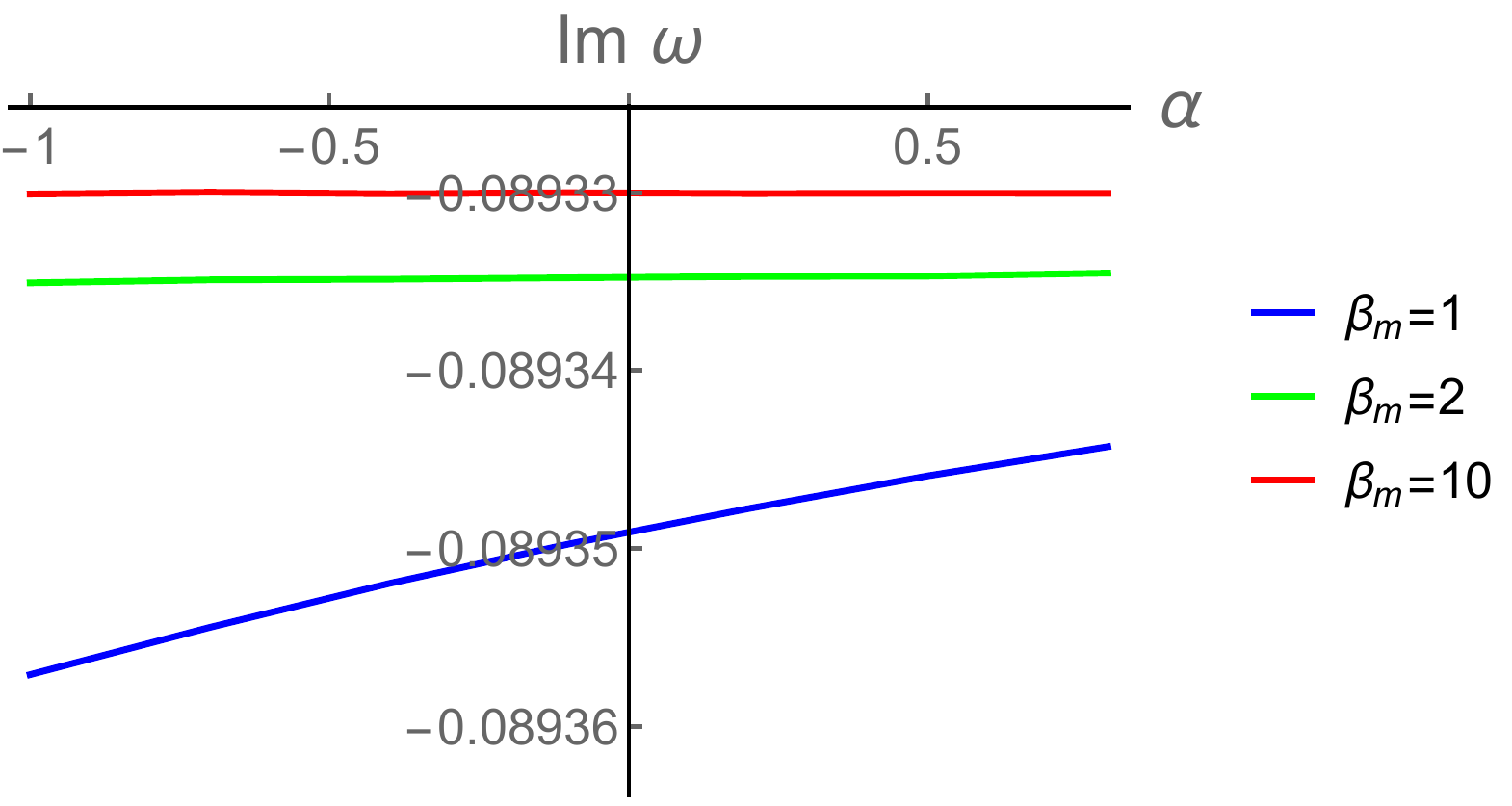}
\caption{The real part (upper) and imaginary part (lower) of the fundamental QNM frequencies of the Born-Infeld black holes in the Palatini $R+\alpha R^2$ gravity are exhibited. The frequencies are shown with respect to $\alpha$ and the results are based on the potential $V_1$ (left) and $V_2$ (right). The multipole number and the charge are fixed to $l=2$ and $Q_*=0.2$, respectively.} 
\label{fig3}
\end{figure*}

For a more general case, we shall consider the Palatini $R+\alpha R^2$ gravity coupled with Born-Infeld NED. In this case, there is no exact expression of the metric functions. The metric functions can only be written in an integral form \cite{Olmo:2011ja}. In Fig.~\ref{fig3}, we rescale $\alpha$ as $\alpha/r_s^2\rightarrow\alpha$, fix the value of the charge $Q_*=0.2$, and exhibit the QNM frequencies with respect to $\alpha$ (we shall mention that the qualitative behaviors of our results remain unchanged when we alter the values of $Q_*$ as long as the charge is smaller than its extremal value). It can be seen that when $\beta_m$ gets large, the frequencies remain almost constant when changing $\alpha$. This is expectable because in this case, the NED reduces to linear Maxwell fields and the Palatini $R+\alpha R^2$ reduces to GR in absence of a cosmological constant. Therefore, the QNMs reduce to those of the RN black hole.

\subsection{Eikonal QNMs}\label{sec.eikonal}
In Ref.~\cite{Cardoso:2008bp}, it has been shown that in GR the QNMs in the eikonal limit ($l\approx\mu\rightarrow\infty$) of any stationary, spherically symmetric, and asymptotically flat black hole can be deduced from the properties of the unstable null circular orbit around the black hole. More precisely, the QNM frequency in the eikonal limit can be expressed as \cite{Cardoso:2008bp}
\begin{equation}
\omega\approx\Omega_cl-i(n+1/2)|\lambda_c|\,,\label{eikonalGRfre}
\end{equation}
where $\Omega_c$ is interpreted as the angular velocity of the null circular orbit and the parameter $\lambda_c$ is the Lyapunov exponent quantifying the instability of the orbit. The derivation of Eq.~\eqref{eikonalGRfre} is related to the fact that for these black holes, the potentials in the master equations within the eikonal limit can be approximated as
\begin{equation}
V=\frac{e^{2\nu}}{r^2}l^2\,.\label{VeikonalGR}
\end{equation}
It can be proven that the peak of this potential \eqref{VeikonalGR} coincides with the radius of the null circular orbit. After inserting the potential \eqref{VeikonalGR} into the the 1st order WKB formula, we can derive Eq.~\eqref{eikonalGRfre}. It can be shown that this equation is valid as well in some modified theories of gravity. In fact, it can be seen from Ref.~\cite{Chen:2018mkf}, and from the master equations \eqref{fr222} and \eqref{fr111} that the massless scalar field perturbations and the axial perturbations of a charged black hole in the Palatini $f(R)$ gravity coupled with NED satisfy the approximation \eqref{eikonalGRfre}. The same is also valid for the massless scalar field perturbations of an EiBI charged black hole \cite{Chen:2018mkf}.

However, Eq.~\eqref{eikonalGRfre} may not be valid for the axial perturbations of the EiBI charged black holes. According to the master equations \eqref{H1EIBIEQ} and \eqref{H2EIBIEQ}, the potentials in the eikonal limit are
\begin{align}
V_1\approx\frac{e^{2\nu}}{r^2}\left(\frac{\beta_g^2r^4+Q_*^2}{\beta_g^2r^4-Q_*^2}\right)l^2\,,\label{eikonalV1}\\
V_2\approx\frac{e^{2\nu}}{r^2}\left(\frac{\beta_g^2r^4-Q_*^2}{\beta_g^2r^4+Q_*^2}\right)l^2\,.\label{eikonalV2}
\end{align}
Note that these approximated expressions are valid for both $\epsilon=\pm1$. Because of the factors $\beta_g^2r^4\pm Q^2$, the relation between the eikonal QNMs and the properties of the null circular orbit around the black hole would be violated. Instead, the QNMs of the axial perturbations described by the potentials \eqref{eikonalV1} and \eqref{eikonalV2} can be expressed as
\begin{equation}
\omega_i\approx\sqrt{V_{i\,p}}-i\left(n+1/2\right)\sqrt{\frac{-V_{i\,p}''}{2V_{i\,p}}}\,,
\end{equation}
where $i=1,2$ and the index $p$ denotes the quantities calculated at the peak of the potentials. Note that at the location of the peak, we have
\begin{equation}
V_{i\,p}'=0\,.
\end{equation}

\begin{figure}[t]
\centering
\graphicspath{{fig/}}
\includegraphics[scale=0.51]{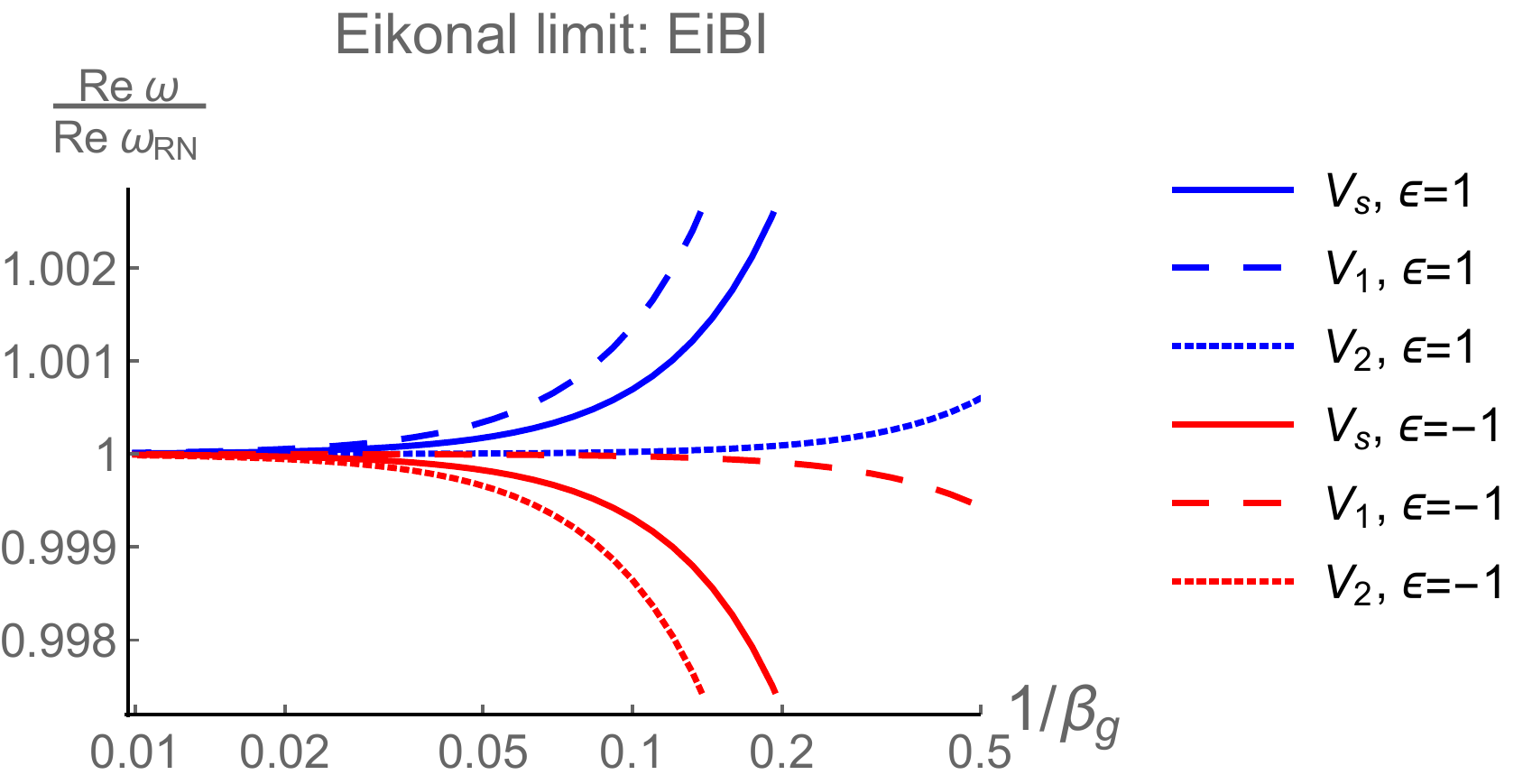}
\includegraphics[scale=0.51]{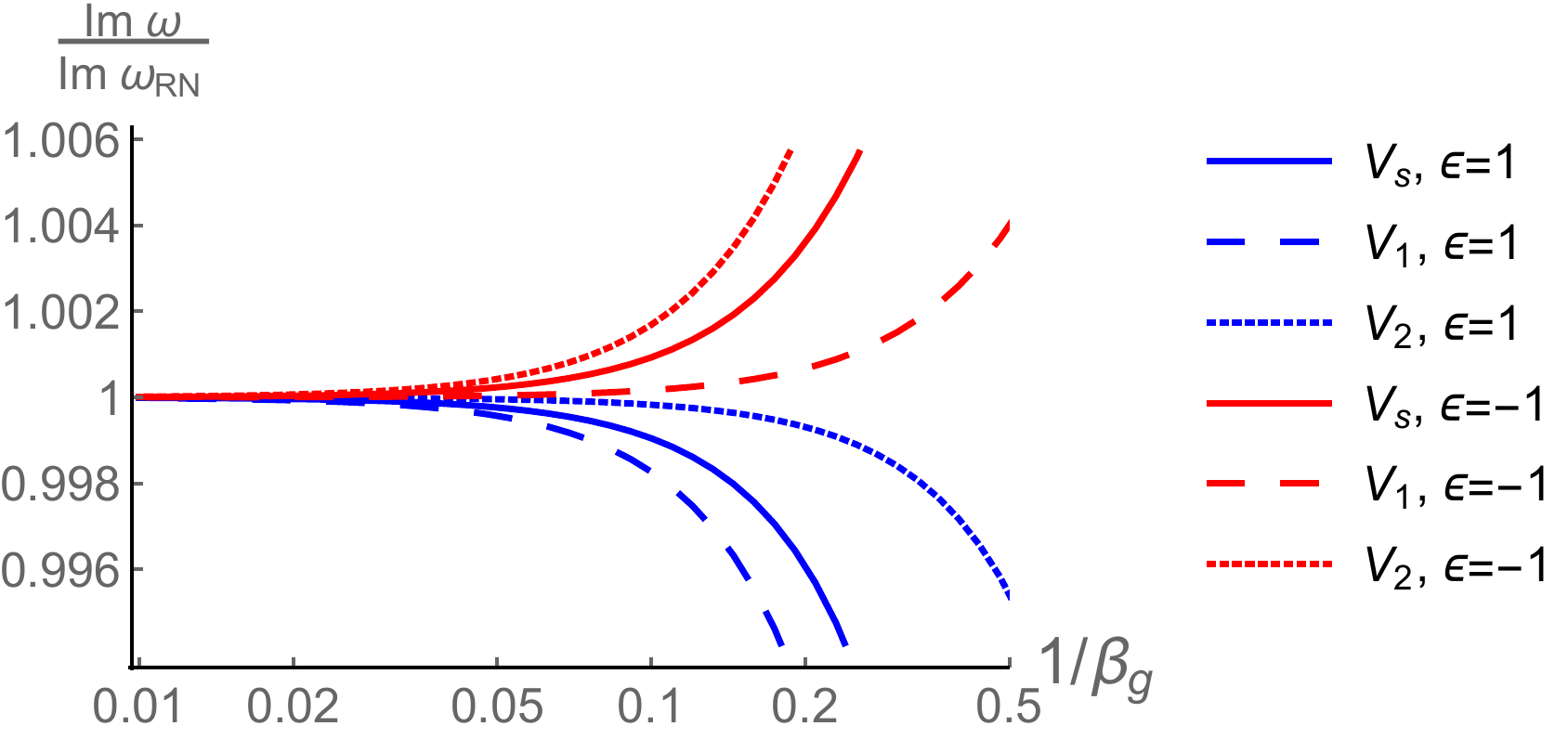}
\caption{The real part (upper) and imaginary part (lower) of the QNMs of the EiBI charged black holes in the eikonal limit ($l\rightarrow\infty$) are shown with respect to $1/\beta_g$. The solid curves are included as well, indicating the QNMs of the massless scalar field perturbations which are described by the potential $V_s=V$ given in Eq.~\eqref{VeikonalGR}.} 
\label{figeikonaleibi}
\end{figure}

In Fig.~\ref{figeikonaleibi}, we exhibit the eikonal QNM frequencies of the EiBI charged black holes in terms of $1/\beta_g$. The charge is fixed to $Q_*=0.4$ (we choose this value to amplify the effects of the charge on the QNMs). The blue (red) curves correspond to a positive (negative) EiBI coupling constant. The dashed and the dotted curves are the eikonal QNMs described by the potential $V_1$ and $V_2$, respectively. We also present the eikonal QNMs for the massless scalar field perturbations which can be described by the potential \eqref{VeikonalGR} in the sense that $V_s=V$ (see Ref.~\cite{Chen:2018mkf}). It can be seen that the eikonal QNMs of the axial perturbations for the EiBI charged black holes (dashed and dotted curves) deviate from those corresponding to the unstable null circular orbit (solid curves). 

Before closing this subsection, we would like to mention that the violation of Eq.~\eqref{eikonalGRfre} for the axial perturbations of the EiBI charged black holes could be due to the non-trivial coupling between the electromagnetic and the gravitational fields in this theory. On the other hand, if we assume that the electromagnetic perturbations do not alter the spacetime geometry, the electromagnetic perturbations will be described by the master equation \eqref{linearcharge} without the metric perturbation terms on the right hand side. In this regard, the potential describing the electromagnetic perturbations in the eikonal limit can be approximated as Eq.~\eqref{VeikonalGR}, and the correspondence proposed in Ref.~\cite{Cardoso:2008bp}, i.e., Eq.~\eqref{eikonalGRfre} is satisfied.

\section{Conclusions}\label{conclu}
In this paper, we consider specifically two gravitational theories within the Palatini formulation and study the QNMs of the axial perturbations for the charged black holes in these theories. These theories of gravity are, respectively, the Palatini $f(R)$ gravity coupled with Born-Infeld NED and the EiBI gravity coupled with linear electromagnetic fields. One of our goals is to see how the Born-Infeld structures from the gravitational sector and from the matter sector change differently the QNM frequencies. Therefore, we pay special attention to the comparison between the QNMs of the EBI black holes and the EiBI charged black holes. The QNMs of the Born-Infeld charged black holes in the Palatini $R+\alpha R^2$ gravity are also discussed. In fact, our paper can be regarded as a further extension of our previous work \cite{Chen:2018mkf} in which we studied the QNMs of the massless scalar field perturbations to these different charged black holes.

By using the tetrad formalism, we have derived the coupled master equations describing the axial perturbations of the charged black holes. In the two theories that we are considering, the coupled equations reduce to those of the RN black holes when the ratio of the charge and the Born-Infeld coupling constant $Q_*/\beta_m$ (or $Q_*/\beta_g$) is small. The QNM frequencies of the charged black holes are evaluated by using the WKB method up to the 6th order, which is accurate for modes whose multiple is larger than the overtone $l>n$. In this paper, we mainly focus on the QNMs of the fundamental modes ($n=0$), since these modes have the longest decay time and would dominate the late time ringdown signals from an astrophysical perspective. Our results indicate that the charged black holes are all stable against the axial perturbations. Besides, the QNM frequencies would deviate from those of the RN black hole when nonlinearity of matter fields (Born-Infeld NED) or modification of the gravitational theory (EiBI or $f(R)$) are considered. For instance, both the absolute value of the real part and the imaginary part of the QNM frequencies for the EBI charged black holes would increase with the value of $1/\beta_m$.  On the other hand, we show that by increasing the value of $1/\beta_g$, the real part of the QNM frequencies and the decay time ($\propto 1/|\textrm{Im}\,\omega|$) would increase (decrease) for the EiBI charged black holes with $\epsilon=+1$ ($\epsilon=-1$).

Furthermore, we study the QNMs of these black holes in the eikonal limit ($l\rightarrow\infty$). Interestingly, we find that the QNM frequencies in this limit for the EiBI charged black holes cannot be described by the properties of the unstable null circular orbit around the black hole. In other words, the QNM formula \eqref{eikonalGRfre} proposed in Ref.~\cite{Cardoso:2008bp} is not valid for the EiBI charged black holes. This violation could be an extra possible imprint from the EiBI corrections on the QNMs, aside from the QNM spectra, that may be detectable in the future.

In addition to the axial perturbations, it is necessary to investigate the QNMs of the polar perturbations (even parity perturbations) for the charged black holes considered in this work. For the Schwarzschild \cite{Zerilli:1971wd} and the RN charged black holes \cite{Chandrabook} in GR, it is well-known that their axial and polar perturbations are isospectral. This means that the potential terms in their master equations satisfy a certain relation in such a way that the QNMs of the axial and polar perturbations have identical spectra. The isospectrality could be violated in the presence of, for instance, nonlinearity in the matter source \cite{Toshmatov:2018ell,Toshmatov:2018tyo}, or modifications of the Einstein-Hilbert action \cite{Bhattacharyya:2017tyc}, and so on. The violation/fulfillment of the isospectrality for the charged black holes in the Palatini-type gravity theories could be an additional tool to test the underlying theories and we shall leave this interesting issue for a coming work.

\appendix 

\section{The decoupling of the master equations}\label{deapp}
\begin{figure*}[tt]
\centering
\graphicspath{{fig/}}
\includegraphics[scale=0.55]{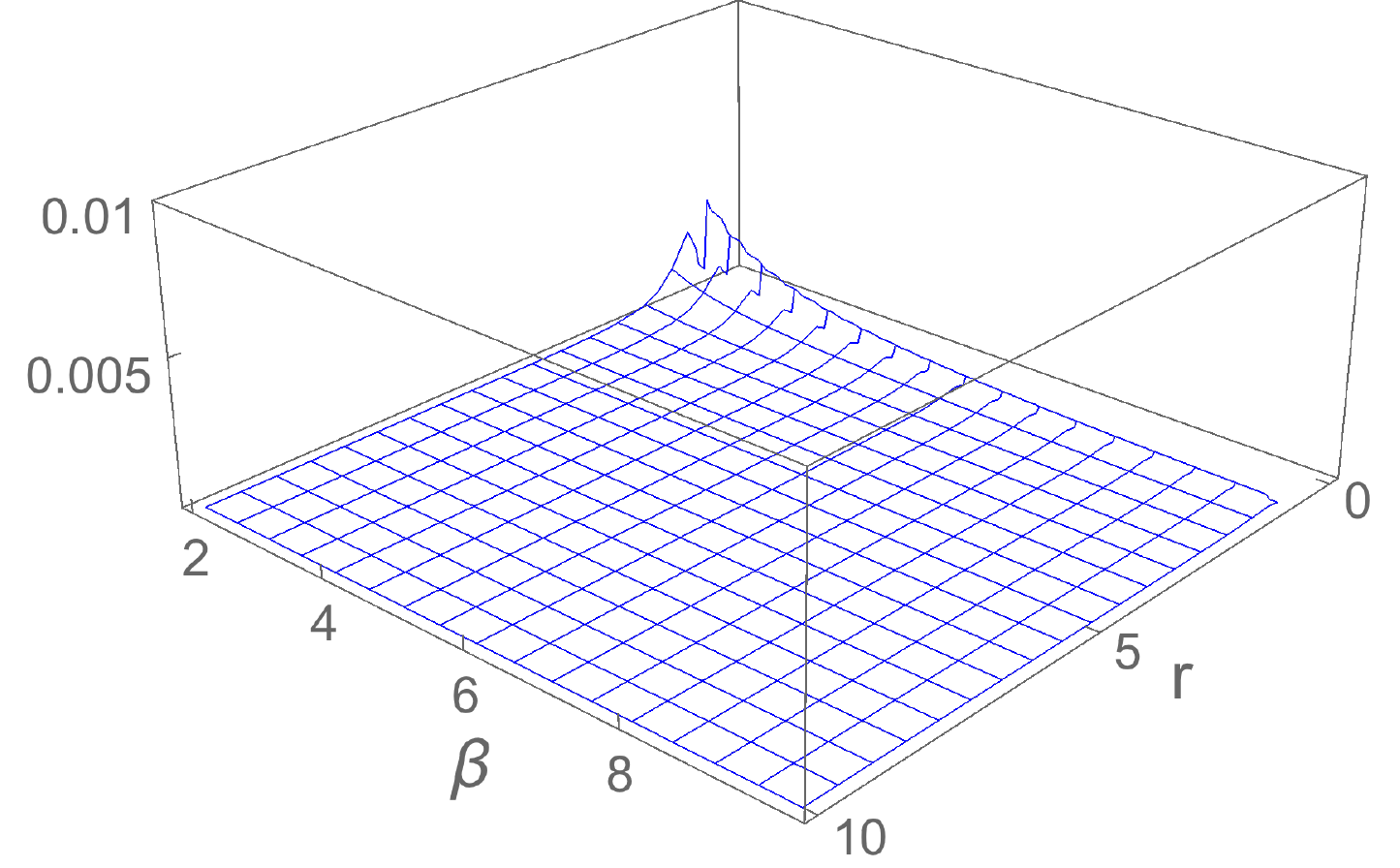}
\includegraphics[scale=0.55]{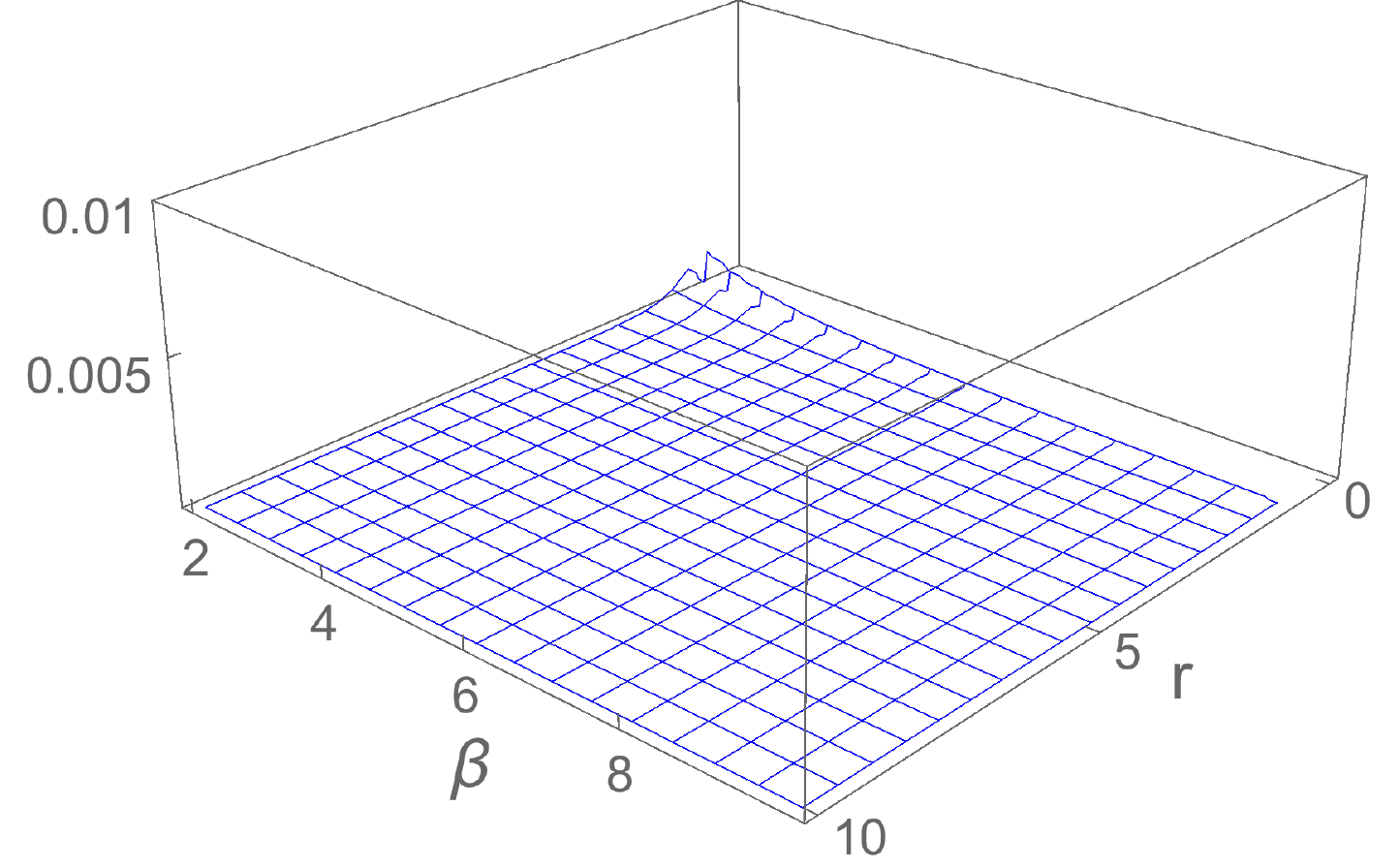}
\caption{The ratio $|\partial_{r_*}\bold{P}_{\textrm{c}}|/|\bold{P}_{\textrm{RN}}|$ for the two eigenvectors of the matrix $\bold{V}$ for the EBI charged black holes is presented with respect to $r$ and $\beta$. We assume here $Q_*=0.3$ and $l=2$. Those values are chosen to estimate the largest errors in our calculations. Note that the domain of $r$ we are considering here is $r\ge r_H$, where $r_H$ is the horizon satisfying $e^{2\nu}=0$.} 
\label{eivenv}
\end{figure*}

It is well-known that for the RN black hole, the coupled master equation \eqref{coupledeq1}, or equivalently
\begin{equation}
\left(\partial_{r_*}^2+\omega^2\right)\bold{H}^{(-)}=\bold{V}\bold{H}^{(-)}\,,\label{matrixbold}
\end{equation}
can be decoupled by diagonalizing the matrix $\bold{V}$ to obtain its eigenvalues $V_1$ and $V_2$: 
\begin{align}
V_1&=\frac{1}{2}\left[V_{11}+V_{22}+\sqrt{\left(V_{11}-V_{22}\right)^2+4V_{12}V_{21}}\right]\,,\nonumber\\
V_2&=\frac{1}{2}\left[V_{11}+V_{22}-\sqrt{\left(V_{11}-V_{22}\right)^2+4V_{12}V_{21}}\right]\,.\nonumber\\\label{diagonizingmatrixapp}
\end{align}
This is because the eigenvectors $\bold{P}=\bold{P}_{\textrm{RN}}$ of the matrix $\bold{V}$ are non-vanishing constant vectors for the RN black hole. Therefore, one can use a similarity transformation $\bold{H}^{(-)}=\bold{P}\bold{\bar{H}}^{(-)}$ to rewrite the matrix equation \eqref{matrixbold} as follows
\begin{equation}
\bold{P}\left(\partial_{r_*}^2+\omega^2\right)\bold{\bar{H}}^{(-)}=\bold{V}\bold{P}\bold{\bar{H}}^{(-)}\,.\nonumber
\end{equation}
By multiplying the above equation by $\bold{P}^{-1}$, the equation can be decoupled.

On the other hand, for the extended charged black holes considered in this paper, the eigenvectors of the matrix $\bold{V}$ are not constant vectors anymore. Instead, they would depend on $r$ and, strictly speaking, one is not able to diagonalize $\bold{V}$ to decouple the master equations. It can be proven that for these extended charged black holes, the eigenvectors can be expressed as
\begin{equation}
\bold{P}=\bold{P}_{\textrm{RN}}+\bold{P}_{\textrm{c}}(r)\,,\nonumber
\end{equation}
where $\bold{P}_{\textrm{c}}(r)$ stands for the correction term. If we use the same similarity transformation $\bold{H}^{(-)}=\bold{P}\bold{\bar{H}}^{(-)}$, the matrix equation \eqref{matrixbold} can be rewritten as
\begin{align}
\bold{P}\left(\partial_{r_*}^2+\omega^2\right)\bold{\bar{H}}^{(-)}&+2\left(\partial_{r_*}\bold{P}_{\textrm{c}}\right)\left(\partial_{r_*}\bold{\bar{H}}^{(-)}\right)+\left(\partial_{r_*}^2\bold{P}_{\textrm{c}}\right)\bold{\bar{H}}^{(-)}\nonumber\\
=&\,\bold{V}\bold{P}\bold{\bar{H}}^{(-)}\,.\label{boldeqnewBH}
\end{align}
In general, Eq.~\eqref{boldeqnewBH} cannot be decoupled due to the last two terms on the left hand side.

However, we will argue that for our present work and for the parameter space of interest, the last two terms on the left hand side of Eq.~\eqref{boldeqnewBH} are actually very small as compared with the other terms. The arguments are the following:
\begin{enumerate}[(i)]
\item In this work, we calculate the QNM frequencies with a semi-analytical approach, which is formulated within the WKB approximation. For the cases where this approach is valid, the wave functions $\bold{\bar{H}}^{(-)}$ can be solved with the WKB approximation. It can then be proven that $\bold{\bar{H}}^{(-)}$ and its derivatives ($\partial_{r_*}\bold{\bar{H}}^{(-)}$ and $\partial_{r_*}^2\bold{\bar{H}}^{(-)}$) have the same order of magnitudes (note that we have normalized all relevant quantities with respect to $r_s$, so the magnitude of the frequencies would be of order one, which is also consistent with our results shown in section~\ref{sectV}).
\item It can be shown that the magnitude of $\partial_{r_*}\bold{P}_{\textrm{c}}$ is very small as compared with the magnitude of $\bold{P}_{\textrm{RN}}$ outside the event horizon and in the parameter space of our interest. In Fig.~\ref{eivenv}, we assume $Q_*=0.3$ and $l=2$, and exhibit the smallness of the ratio $|\partial_{r_*}\bold{P}_{\textrm{c}}|/|\bold{P}_{\textrm{RN}}|$ for the EBI black hole, with respect to $r$ and the value of $\beta$. For the Born-Infeld black holes with $f(R)$ being a quadratic function and for the EiBI charged black holes, this ratio is also very tiny.
\end{enumerate}

According to the arguments above, the last two terms on the left hand side of Eq.~\eqref{boldeqnewBH} are very small as compared to the other terms. Therefore, when studying the QNMs of the extended charged black holes, we shall omit these two terms and decouple the master equation \eqref{coupledeq1} by diagonalizing the potential matrix $\bold{V}$ as given in Eq.~\eqref{diagonizingmatrixapp}.

\acknowledgments

CYC would like to thank R. A. Konoplya for providing the WKB approximation. CYC and PC are supported by Taiwan National Science Council under Project No. NSC 97-2112-M-002-026-MY3, Leung Center for Cosmology and Particle Astrophysics (LeCosPA) of National Taiwan University, and Taiwan National Center for Theoretical Sciences (NCTS). MBL is supported by the Basque Foundation of Science Ikerbasque. She also would like to acknowledge the partial support from the Basque government Grant No. IT956-16 (Spain) and from the project FIS2017-85076-P (MINECO/AEI/FEDER, UE). PC is in addition supported by US Department of Energy under Contract No. DE-AC03-76SF00515. MBL is also thankful to LeCosPA (National Taiwan University, Taipei) for kind hospitality while part of this work was done.

\end{document}